\documentclass[fleqn,usenatbib]{mnras}

\usepackage{newtxtext,newtxmath}
\usepackage{subfloat}
\usepackage{tabularx}
\usepackage{longtable}
\usepackage[T1]{fontenc}
\usepackage{ae,aecompl}

\usepackage{graphicx}	
\usepackage{amsmath}	
\usepackage{subfig}
\usepackage{supertabular,booktabs}
\usepackage{threeparttable}
\usepackage{longtable}
\usepackage{ltcaption}

\usepackage{svg}
\newcommand{\orcid}[2]{\href{https://orcid.org/#1}{\includesvg[width=10pt]{orcid}}}

\title[Stellar Masses of Clump]{Stellar Masses of Clumps in Gas-rich, Turbulent Disk Galaxies}

\author[Ambachew et al.]{
Liyualem Ambachew,$^{1,2}$\thanks{E-mail: ltilahun@swin.edu.au}
Deanne B. Fisher $^{1,2}$,
Karl Glazebrook $^{1,2}$,
Marianne Girard$^{1,2}$,
\newauthor
Danail Obreschkow$^{3,2}$,
Roberto Abraham$^{4}$,
Alberto Bolatto$^{5,6,7}$,
Laura Lenki\'{c}$^{5,9}$ and 
\newauthor
Ivana Damjanov$^{8}$
\\
$^{1}$Centre for Astrophysics and Supercomputing, Swinburne University of Technology, PO Box 218, Hawthorn, VIC 3122, Australia\\
$^{2}$ARC Centre of Excellence for All Sky Astrophysics in 3 Dimensions (ASTRO 3D)\\
$^{3}$International Centre for Radio Astronomy Research (ICRAR), M468, University of Western Australia, 35 Stirling Hwy., Crawley, WA 6009, Australia\\
$^{4}$ Department of Astronomy \& Astrophysics, University of Toronto, 50 St. George St., Toronto, ON M5S 3H8, Canada\\
$^{5}$ Department of Astronomy, University of Maryland, College Park, MD 20742, USA\\
$^{6}$ Visiting Scholar, the Flatiron Institute, Center for Computational Astrophysics, NY 10010, USA\\
$^{7}$ Visiting Astronomer, National Radio Astronomy Observatory, VA 22903, USA\\
$^{8}$ Department of Astronomy \& Physics, Saint Mary's University, 923 Robie St, Halifax, Nova Scotia,  Canada \\
$^{9}$ SOFIA Science Center, USRA, NASA Ames Research Center, M.S. N232-12, Moffett Field, CA 94035, USA
}

\date{Accepted XXX. Received YYY; in original form ZZZ}

\pubyear{2015}

\begin{document}
\label{firstpage}
\pagerange{\pageref{firstpage}--\pageref{lastpage}}
\maketitle

\begin{abstract}
In this paper we use \emph{HST/WFC3} observations of 6  galaxies from the DYNAMO survey, combined with stellar population modelling of the SED, to determine the stellar masses of DYNAMO clumps. The DYNAMO sample has been shown to have properties similar to $z\approx1.5$ turbulent, clumpy disks. DYNAMO sample clump masses offer a useful comparison for studies of $z>1$ in that the galaxies have the same properties, yet the observational biases are significantly different. Using DYNAMO we can more easily probe rest-frame near-IR wavelengths and also probe finer spatial scales. 
We find that the stellar mass of DYNAMO clumps is typically  $10^{7}-10^8 \mathrm{M}_\odot$. We employ a technique that makes non-parametric corrections in removal of light from nearby clumps, and carries out a locally determined disk subtraction. The process of disk subtraction is the dominant effect, and can alter clump masses at the 0.3~dex level.  Using these masses, we investigate the stellar mass function of clumps in DYNAMO galaxies. DYNAMO stellar mass functions follow a declining power law with slope $\alpha \approx -1.4$, which is slightly shallower than, but similar to what is observed in $z>1$ lensed galaxies. We compare DYNAMO clump masses to results of simulations. The masses and galactocentric position of clumps in DYNAMO galaxies are more similar to long-lived clumps in simulations. Similar to recent DYNAMO results on the stellar population gradients, these results are consistent with simulations that do not employ strong ``early'' radiative feedback prescriptions. 


\end{abstract}



\section{Introduction}
\label{sec:introduction}

Most of the stars in the universe are formed at $z \sim 2$ \citep[e.g.,][]{Hopkins2006,Madau2014}. The first Hubble Deep Field surveys revealed that the highly star forming galaxies at this epoch have irregular morphologies compared to local Hubble Sequence galaxies and are dominated by a few bright patches \citep{Abraham_1996}. When these bright patches are viewed edge-on, they appear in  lines, and thus often referred to as 'chain galaxies' \citep{Cowie_1995}. The appearance of these star forming galaxies suggests that these bright knots are embedded in disk-like systems. Follow-up studies showed that they represent massive clusters of young stars with sizes of a $\sim$ kpc  \citep[e.g.,][]{Elmegreen_2005}. These giant clumps are  detected in the deep and high resolution rest-frame UV \citep[e.g.,][]{Elmegreen_2005,Schreiber2009}, as well as in the rest rest-frame optical line emissions \citep[e.g.,][]{Genzel_2011, Jones_2010,Livermore2012,Livermore2015,Fisher_2017}.

Early explanations of origin of clumps focused primarily on major mergers  \citep[e.g.,][]{Conselice_2003}. However, later kinematic studies of star forming galaxies at redshift $z \sim 2$  \citep{Genzel_2006, Shapiro_2008, Schreiber2009} revealed that a significant fraction of clumpy galaxies do not show ongoing merger signatures. Rather the galaxies have rotating velocity fields in spite of their clumpy morphology.

A popular theory for clump formation, currently, is via self-gravitating instabilities \citep[e.g.,][]{Noguchi1999,Bournaud_2007,Elmegreen_2008,Dekel_2009,Ceverino_2010}. The scenario of clumps being formed through violent gravitational instabilities is supported by some observational studies such as  \citet[][]{Bournaud_2007,Genzel_2011,Guo_2012,Fisher_2017,White_2017}. Alternate evidence of clump formation through self-gravitating disk is found in their stellar mass function. If clumps form internally, their resulting stellar mass function would follow a declining power law with a slope of $\sim -2$ \citep[e.g.,][]{Dessauges_2018}. The recent VELA simulation study by \cite{Huertas2020} finds that slope of the clump stellar mass function can be $\sim -1.5$. Alternate theories for how clumps form in disks have been put forward, including through spiral arm instabilities \citep{Inoue2018} or bottom up collisions of smaller clumps \citep{Behrendt_2016}. While it is likely that many clumpy galaxies at $z>1$ are indeed the result of merging, it is nonetheless commonly accepted that some disk processes are generating large star-forming complexes within gas-rich disks. 

Estimating the lifetime of clumps remains a one of the main goals in this field, for multiple reasons. First, it is important to understand the ultimate fate of clumps, and if they contribute to bulges in local Universe. Secondly, a wealth of simulation work suggests that the details of feedback prescriptions have significant impact on lifetime, and therefore current stellar mass of clumps \citep{Hopkins_2012,Mayer_2016,Mandelker_2016}. To first order, this can be understood in that clumps have star formation rate (SFR) surface densities that are  orders-of-magnitude higher than what is observed in local spirals; subtle changes to stellar feedback models can have amplified affects in inside clumps.  

Estimating the longevity of clumps,however, is not a simple task. \cite{Lenkic_2021} show, also with DYNAMO galaxies, that individual clumps do not have a single age, but gradients in color profiles indicate a growing system. These results imply that using a single age for clumps, derived from stellar populations will be biased toward younger ages than the true age of clumps. The simulations of \cite{Bournaud_2014} show that  clumps can continue to accrete gas for new star formation, rendering the light-weighted stellar population a poor metric of the clump lifetime, as it may always be young.  \cite{Lenkic_2021} find that using the maximum age found in clumps improves clump age-galactocentric distance correlations, and may indicate a long-lived system. 



A significant body of work suggests that resolution and sensitivity are critical for measuring the properties of clumps, including their stellar mass. HST observations of unlensed galaxies suggest that stellar masses of clumps are roughly $10^{8}-10^{9}$ $\mathrm{M}_\odot$ \citep[e.g.,][]{Schreiber_2011,Guo_2012}. However,  studies of gravitational lensed high-z systems find a characteristic clump luminosity that is lower than in unlensed studies high-z galaxies \citep[e.g.,][]{Jones_2010,Livermore2012,Dessauges_Zavadsky_2017,Cava_2018}. \cite{Dessauges_Zavadsky_2017} compare stellar masses of clumpy star forming galaxies at redshift of $1.1 <z<3.6$ in lensed or unlensed galaxies. They argue that the systematic uncertainties due to spatial resolution and sensitivity significantly affect the selection and measurement of clump properties, which cause the resulting clump stellar mass distribution to be biased toward higher masses. Moreover, \cite{Cava_2018} analysed multiple images of the same gravitationally lensed galaxy at different magnification, finding that unlensed observations with $\sim1$~kpc resolution tends to overestimate both the size and luminosity of clumps, and therefore the mass. A particular problem with unlensed studies of clumps is the blending of smaller clumps into a single resolution element. \cite{Fisher_2017} show in more detail how the clustering of smaller clumps when imaged at $\sim1$~kpc resolution results in clump properties similar to those in unlensed $z\sim2$. They find that fluxes and sizes are significantly impacted when images are blurred. In addition to observational studies, simulation results also show that $100$~pc resolution is needed to isolate clumps \citep{Tamburello_2017}. 


In this paper, we use a sample of local galaxies called DYnamics of Newly-Assembled Massive Object \citep[DYNAMO;][]{Green_2014}.  The DYNAMO sample was selected from the Sloan Digital Sky Survey to be the brightest H$\alpha$ emitting galaxies that do not contain AGNs. DYNAMO galaxies have properties that are similar to those of star-forming $z\sim 2$ galaxies, but yet are located at $z\sim0.1$ \citep[e.g,][]{Green_2014,Bassett_2014,Oliva_2017,Fisher_2014,White_2017,Fisher_2019,Girard2021}. Therefore they can be observed with significantly improved spatial resolution and deeper sensitivity, and allow us to measure clump properties with greater precision addressing the biases discussed above.


In this work we will use results from HST observations, combined with stellar population modelling of the spectral energy distribution (SED), to estimate the masses of individual clumps in DYNAMO galaxies. Our results allow for a controlled comparison of clumps with improved spatial resolution and wavelength coverage for comparison to results of large surveys of $z\sim2$ observations with Hubble Space Telescope (HST) and in the near future James Webb Space Telescope (JWST).


\section{The Data and Sample}
\subsection{Sample}
For this work, we used a subset of six galaxies from the DYNAMO sample. The full DYNAMO sample is described in detail in \cite{Green_2014}.  DYNAMO galaxies were selected from Sloan Digital Sky Survey (SDSS) DR4 \citep{Adelman_2006}, excluding galaxies with AGNs. DYNAMO galaxies are found in the local universe within redshift range of $z=0.075-0.13$ that have a high star formation rate.  The full DYNAMO sample was then observed with integral field spectroscopy observations of H$\alpha$. 

In this study we use a subset of more well studied galaxies from DYNAMO. The key feature of the follow-up DYNAMO sub-sample is that a kinematic selection was made from the original DYNAMO survey \citep{Green_2014}. The DYNAMO team identified a sub-sample of galaxies with rotating kinematics and high velocity dispersion in the seeing limited observations. Those galaxies were then rigorously tested to confirm their similarity to $z\approx1-2$ galaxies. After those test we have then concentrated on those targets, which more robustly match the observations of distant galaxies. 

Several studies show similarities between DYNAMO galaxy properties and $z\sim 1.5$ star forming galaxies.To investigate the morphological similarity between DYNAMO and z ~ 2 galaxies, \cite{Fisher_2017} degrade the resolution of \emph{HST} H$\alpha$ maps of DYNAMO galaxies to the physical resolution of $z\sim 2$ galaxies. They found, at matched resolution, that the clumps in DYNAMO galaxies are as bright as and have similar diameters to $z\sim 2$ galaxies. Clumps in DYNAMO galaxies also meet the definition of clumpy galaxies as defined by the CANDELS results \citep{Guo_2015}. DYNAMO galaxies have high internal gas velocity dispersions, $\sigma\approx 20-100$~km~s$^{-1}$ \citep{Green_2014,Bassett_2014,Oliva_2017,Girard2021}. They also have lower angular momentum for their stellar mass \citep{Obreschkow_2015}, which is consistent with  $z\sim 2$ star forming galaxies with similar mass \citep{Swinbank_2017}. Galaxies at $z\sim 1-2$ have been shown to have gas fractions of $f_{gas}$ $\sim 20-80 \%$ \citep{Tacconi_2013}. However, local spirals have gas fractions roughly $\sim 1-5 \%$ \citep{Saintonge_2012}. DYNAMO galaxies have been shown to have high fractions of molecular gas, $M_{gas}/M_{tot}\approx 10-70$\% , $f_{gas}$ like $z\sim 2$ galaxies \citep{Fisher_2014,White_2017,Fisher_2019}; and have large clumps in both H$\alpha$ gas and restframe $UV$ starlight that are consistent with large clumps of star formation similar to what is observed at $z>1$ \citep{Fisher_2017,Lenkic_2021}. In the most recent DYNAMO study, \cite{Lenkic_2021} shows that these clumps are not likely the result of ``holes" in extinction profiles, but are structures of stars. This has also been shown  by \citet{Elmegreen_2005} in  $z\sim 2$ galaxies, where they find clumps have high average density $\sim 0.2~ \mathrm{M}_\odot$ $pc^{-3}$.

A key issue with studies of highly star forming galaxies in the local Universe is identifying merging galaxies. For DYNAMO, we use three indicators associated to merging. First, we observe kinematics with higher spatial (0.15-0.7 arcsec) resolution  \citep{Bassett_2014,Oliva_2017}, second we observe dust temperatures \citep{White_2017}; and finally we use HST-based high spatial resolution maps of HST~H$\alpha$ \& 600~nm continuum \citep{Fisher_2017}. We only selected systems that are not consistent with merging, as indicated by order rotation in higher spatial resolution ionized gas velocity maps, lower dust temperatures ($T_{dust}\sim20-30$~K) and exponentially decaying stellar surface brightness profile \citep{Fisher_2017}.

Generally, the properties of DYNAMO galaxies resemble galaxies at $z\sim 2.$ DYNAMO galaxies are therefore used as laboratories for studying the processes in clumpy, turbulent disks with higher resolution and sensitivity. Our sample properties spans stellar mass $(1.7-6.4)\, \times\, 10^{10}$\, $\mathrm{M}_\odot$, star formation rate $\mathrm{SFR} \sim 6.9-21\, \mathrm{M}_\odot$ $\mathrm{yr}^{-1}$ and extinction A(H$\alpha$) $\sim 0.59-1.27$ mag. The properties of these six DYNAMO galaxies are listed in Table \ref{tab:sample_table}.

\begin{table*}
	\centering
	\caption{Properties of six DYNAMO - \emph{HST} Targets.}
	\label{tab:sample_table}
	\begin{threeparttable} 
\begin{tabular}{lccccccccc} 
        \toprule
Galaxy & z &  RA &Dec& SFR$^{a}$& $M^a_{*}$&$f^b_{gas}$& $\sigma^b$ &$V^b_{flat}$&$\mathrm{A_{H_{\alpha}}}${$^{c}$}\\
		 & & & &[$\mathrm{M}_\odot$ $yr^{-1}$] &[$10^{10}$ $ \mathrm{M}_\odot$] &&[$\mathrm{kms^{-1}}$]& [$\mathrm{kms^{-1}}$]&[mag] \\ 
		
		\hline
		G04-1 & 0.1298 & 04:12:19.7100 & -05:54:48.60&41.6$\pm2.2$ &6.45&0.33$\pm$ 0.04& 50& 269&1.52 $\pm$ 0.26\\
		G14-1 & 0.1323 &14:54:28.3300 & +00:44:34.30&8.3$\pm0.9$ &2.23&0.77 $\pm$ 0.08 & 70&136&...\\
		G08-5 & 0.1322 &08:54:18.7300 & +06:46:20.60&16.6$\pm1.0$ &1.73& 0.30 $\pm$ 0.05&64& 243&...\\
		G20-2 & 0.1411 &20:44:2.9150 &-06:46:57.90&17.3$\pm0.7$ &2.16&0.21 $\pm$ 0.05&81&166&0.89 $\pm$ 0.1\\
		D13-5 & 0.0753 & 13:30:07.009 & 00:31:53.450 &21.2$\pm0.9$ &5.38&0.36 $\pm$ 0.02&46&192 &1.80 $\pm$ 0.52\\
		D15-3 & 0.0671 &15:34:35.3900&-00:28:44.50&13.7$\pm1.0$ &5.42&0.17$\pm$ 0.04&45&240&...\\
		\bottomrule
		\end{tabular}
		\begin{tablenotes}
        \item[a]{Values from \citet{Green_2014}} 
        \item[b]{Values from \citet{Fisher_2017,Fisher_2019}} \item[c]{H$\alpha$ extinction from \citet{Bassett2017}}
        \end{tablenotes}
        \end{threeparttable} 
	
\end{table*}

\begin{figure*}
	\includegraphics[width=\textwidth]{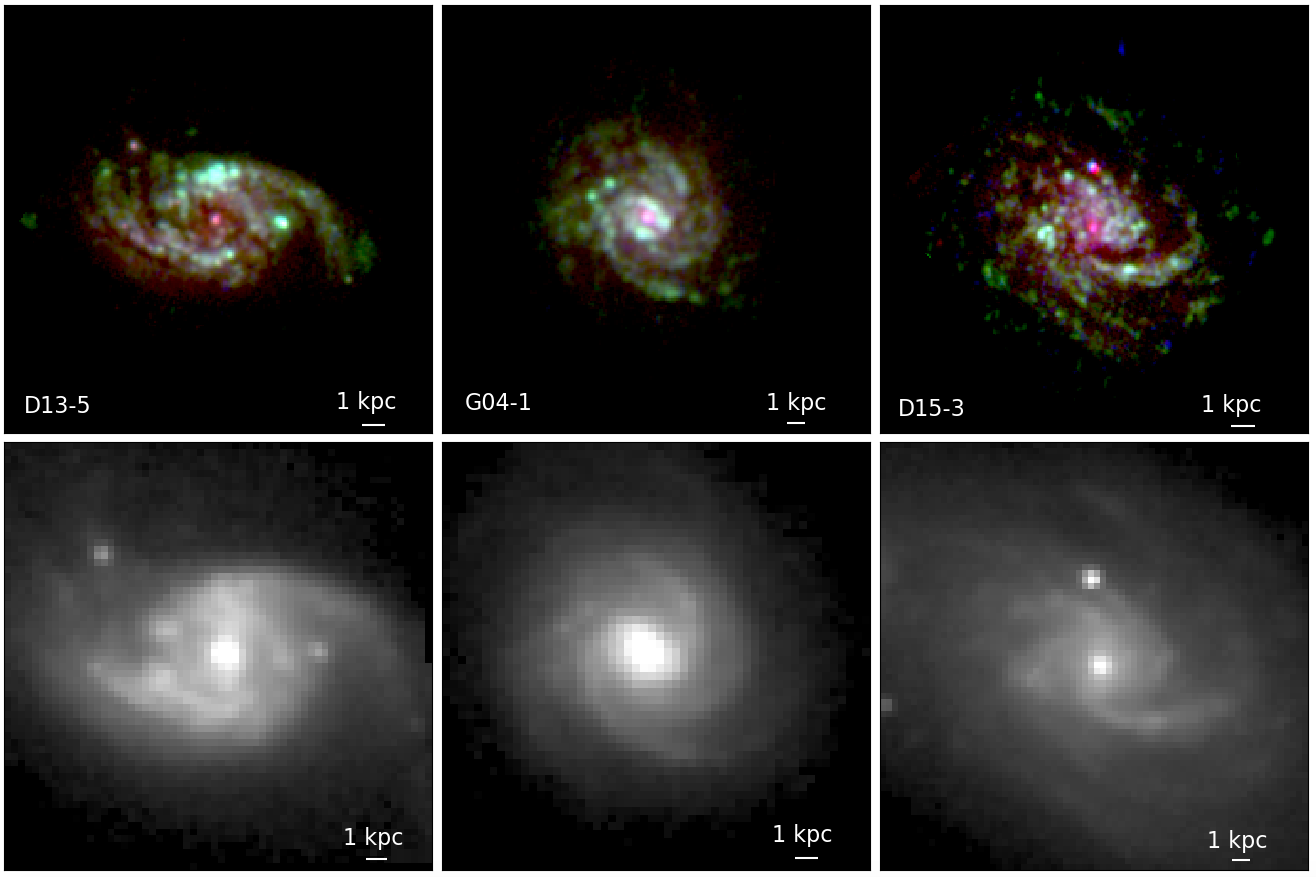}
  \caption{Three color images of DYNAMO galaxies D13-5, G04-1 and D15-3 created using F336W, F467M and FR647M filters at $\sim$0.05'' resolution (top row). Bottom row shows the same galaxies observed at $\sim$0.13'' resolution in the F125W filter. The white line in the bottom right corner of each image corresponds to 1 kpc. Though clumps are less prominent in the F125W image, we do still see large knots of emission.}
   \label{fig:obser_image1}
\end{figure*}

\begin{figure*}
	\includegraphics[width=\textwidth]{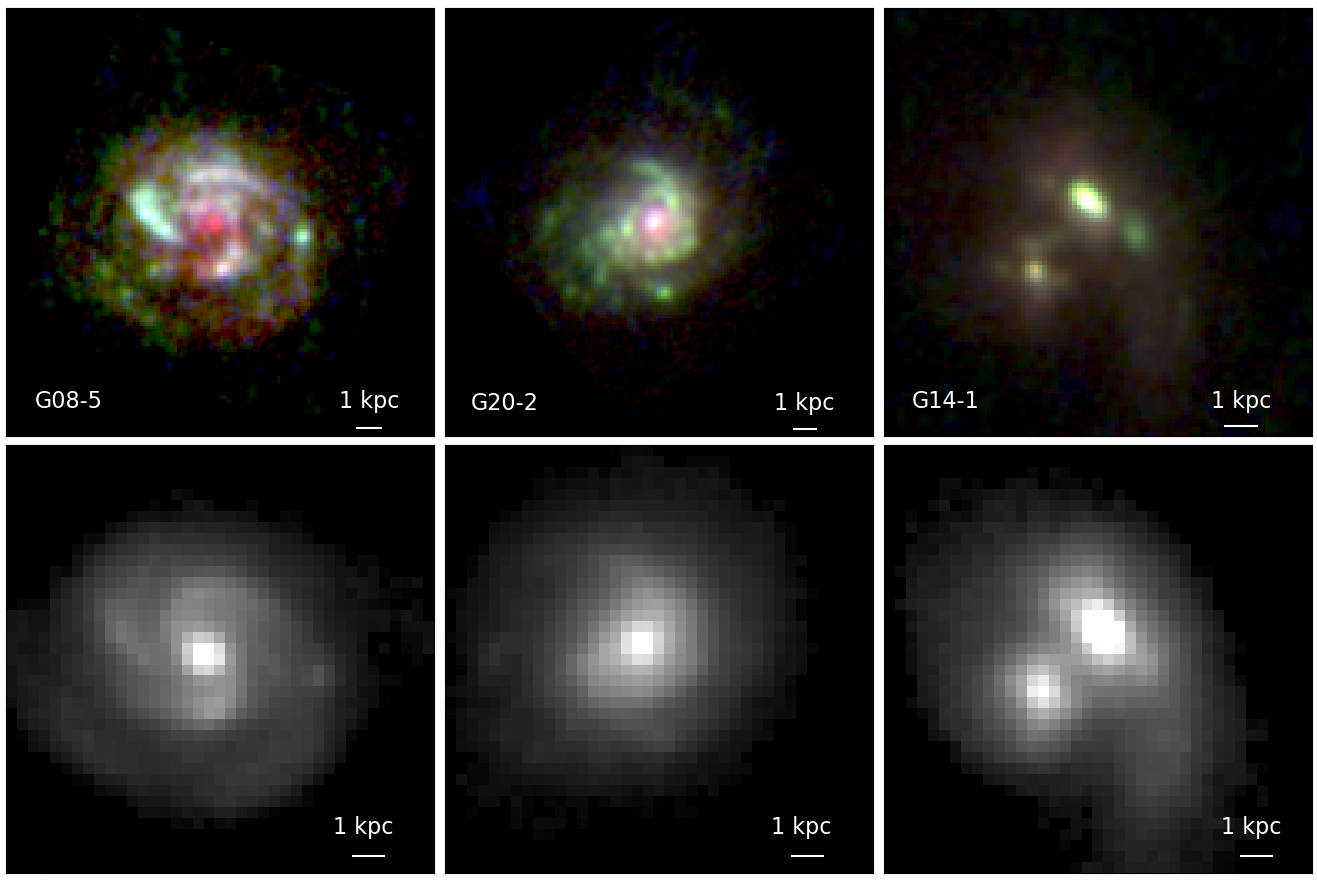}
   \caption{The same as Figure ~\ref{fig:obser_image1} but for target G20-2, G14-1 and G08-5, respectively}
   \label{fig:obser_image2}
\end{figure*}

\subsection{DYNAMO Galaxies as Clumpy Disk}
A number of works in the literature establish the similarity of DYNAMO galaxies to turbulent, gas rich disk galaxies, more commonly observed at $z\sim 1-3$ \citep[e.g][]{Green_2014,Fisher_2014,Bassett_2014,Obreschkow_2015,Fisher_2017,Fisher_2017b,Bassett2017,Oliva_2017,White_2017,Fisher_2019,Girard2021,Lenkic_2021}. \cite{Green_2014} shows that in the overall sample, $\sim$84\% of DYNAMO galaxies have disk-like light profiles and $\sim 50\%$ are located on the Tully-Fisher relation \citep{Green_2014}. Follow-up observations of DYNAMO galaxies were made a sample down selected to be more consistent with having rotating disk kinematics, high SFR, and high H$\alpha$ velocity dispersion. \cite{Bassett_2014} used deep Gemini/GMOS observations to show that DYNAMO galaxies are rotating in both stars and gas. \cite{Oliva_2017} used Keck adaptive optics observations to show that the observation of rotation is not an artifact of poor resolution, but still holds with resolution of $\sim$150~pc. \cite{Fisher_2017b} found that from 10 DYNAMO galaxies, 8 (6 are analyzed in this work) of them have  disk-like exponential decaying surface brightness profile using HST $600$~nm continuum maps. \cite{Fisher_2017b} shows using HST images that the H$\alpha$ emission is ``clumpy". They find that when you degrade the DYNAMO images to match the same physical resolution and sensitivity as found in AO images of $z\approx 1.5$ galaxies that the distribution of sizes and luminosity of detected clumps is similar to those in high-$z$ samples \citep[e.g.][]{Genzel_2011,Guo_2015}. Both \cite{Fisher_2017} and \cite{White_2017} show that the gas densities and kinematics of DYNAMO disks are consitent with marginal stability, i.e. $Q\approx1$. This is consistent with standard expectations of high-$z$ disks. 

There are a several papers outlining the star formation and gas properties of DYNAMO galaxies. A key difference from local galaxies is the gas fraction. DYNAMO galaxies have $f_{gas}\approx 0.15-0.6$ \citep{Fisher_2014,White_2017,Fisher_2019,Girard2021}, which is similar to $z\approx1.5$ galaxies, and makes them 99$^{th}$ percentile outliers from samples of local rotating galaxies of similar mass \citep{Tacconi_2018,Saintonge_2012}. The dust in DYNAMO galaxies is found to have low dust temperatures, $T_{dust}\approx20-30$~K \citep{White_2017}, which is dissimilar from local Universe galaxies of the same infrared luminosity. The low dust temperature implies the geometry of the gas and dust is more like that of thick-disks, as in $z\sim 1-2$ galaxies. Similarly, work in progress (Lenki\'{c} et al. {\em in prep}) finds that the CO line ratios are similar to those of BzK galaxies \citep{Daddi2015}. \citep{Fisher_2019} reports the depletion time of DYNAMO galaxies to be anti-correlated  with the gas velocity dispersion. The clumpy, high velocity dispersion DYNAMO galaxies have depletion times of order $t_{dep}\approx 0.3-1.0$~Gyr, which is similar to observations of $z\approx1.5$ main-sequence galaxies \citep{Tacconi_2018}. Systems with less prominent clumps and lower velocity dispersion (e.g. D15-3 in this work) have $t_{dep}\approx 1-2$~Gyr. The DYNAMO team is currently working on producing a kiloparsec-scale resolved Kennicutt-Schmidt analysis, which includes galaxies from this sample. DYNAMO galaxies are found to have high molecular gas surface densities ($\Sigma_{mol}\approx 10^2-10^3$~M$_{\odot}$~pc$^{-2}$). We refer the reader to Lenki\'{c} et al. ({\em in prep}) for those results, see also the recent work \cite{Fisher2022}.

The origin of DYNAMO galaxies remains a mystery. The gas fractions of DYNAMO galaxies imply that they must have recently undergone an event that mimicked a gas-rich accretion event. However, simply having large gas reservoirs is not sufficient to lead to clumpy star formation and turbulent kinematics. \cite{Catinella2015} find a sample of very gas rich disks at similar redshift; these disks are selected for HI brightness and are not compact like DYNAMO. They do not find high gas velocity dispersions, nor clumpiness \citep{Cortese2017}. The DYNAMO team is currently working to estimate the total gas reservoir of DYNAMO galaxies (Obreschkow et al {\em in prep}).  One observation that is particularly informative is that they are low angular momentum systems. \cite{Obreschkow_2015} compares DYNAMO galaxies to the location in the $j_*-M_*$ diagram. DYNAMO galaxies-- despite being rotating, exponential disks-- are found to have low angular momentum, more similar to galaxies at $z\approx1-2$ \citep{Swinbank_2017,espejo2022}. The implication is that in order to be a clumpy galaxy both large gas accretion and low angular momentum are necessary.


\subsection{DYNAMO \emph{HST} Observation}
Observations of our six DYNAMO galaxies were taken during the Hubble Space Telescope (\emph{HST}) Cycle 25 program (Proposal ID 15069, PI: D.B.Fisher) using WFC3 UVIS and IR modes. These observations were performed using UV band-pass (F225W), optical band-passes (F336W, F467M) and near-IR band-pass (F125W) filters, as well as FR647m filter from an ACS/WFC Cycle 20 program (Proposal ID 12977, PI: I. Damjanov). 

Our data set covers a  wavelength range from  near-UV to  near-IR. This is intended to reduce well-known degeneracies between mass-to-light ratio, extinction and metallicity \citep{Bell_2001}. Our principle aim is to use spectral energy distribution fitting techniques to measure the stellar mass and mass-to-light ratio in sub-kpc regions in clumpy galaxies. \cite{Taylor2011} show that of the stellar population parameters derived from SED fitting, mass-to-light ratio is relatively robust, with more significant degeneracies occurring among other parameters (e.g., age, extinction, metallicity). Previous studies show that using optical-plus-near-IR data significantly increases the ability to derive robust mass-to-light ratios from SED fitting methods \citep[e.g.,][]{Bell_2001,Zibetti_2009}. This principle motivates our choice of five filters ranging from rest-frame $\sim$200~nm to $\sim$1100~nm. In comparison to samples of galaxies at $z>1$, the DYNAMO sample is the only set of clumpy disk galaxies in which rest-frame near-IR observations are possible. We directly assess the effect that the inclusion of near-IR data set has on the derived clump stellar mass for our DYNAMO sample in Section \ref{sec:mass_to_light_ratio} and Figure~\ref{fig:m2l_filters}.

We also have chosen optical filters to avoid strong emission lines (e.g. H$\beta$, [OIII] 5007, and H$\alpha$) in our target galaxies. We present HST composite image created from F336W, F467M and FR647M filters and single-filter F125W images Figures~\ref{fig:obser_image1} and ~\ref{fig:obser_image2}. The top panel shows composit RGB image and the bottom panel shows the same galaxies in F125W filters. All images were reduced using the standard HST pipeline and combined with \textsc{DRIZZLE}.

\section{Methods}
\label{sec:methods} 
In the following subsections we give detailed descriptions of each step in our process. First, we give a general overview.

In order to measure the masses of individual clumps we must first locate clumps, then we used SED fitting code to measure the clump stellar mass. Clump positions are located in the F336W images, which is similar to clump position selection in $z\approx1-3$ HST surveys \citep[e.g.,][]{Guo_2015}.

The spatial resolution of the WFC3/IR instrument used to measure the F125W band images is 0.13'' per pixel. This resolution is more than a factor of 2$\times$ larger than the resolution of WFC3/UVIS and ACS/WFC images that cover the optical light. Moreover, based on results from \cite{Fisher_2017}, it leaves clumps poorly resolved with only 1-2 WFC3/IR resolution elements per clump. We therefore test the systematic bias of using or omitting F125W on the derived mass-to-light ratios. If derived mass-to-light ratios without F125W do not differ significantly from those with F125W we will then opt for the higher spatial resolution measurements. 

We will therefore consider two separate resolutions for our mass measurements. First, we convolve and resample all bands to match the spatial resolution of F125W. We then measure the mass-to-light ratio in each resolution element within the galaxy, as described in section \ref{sec:mass_to_light_ratio}. Second, we convolve and resample all bands, except F125W, to match the spatial resolution of FR647M.

\subsection{PSF Convolution}
\label{sec:psf_convolution}
In order to generate matched point spread function (PSF) sets of images we use standard IRAF packages. We created two sets of images. The first set is matched to the UVIS/IR F125W resolution, which has FWHM  $\sim$0.13''. This set of images was used to measure the mass-to-light ratio (from all bands) in each resolution element within the galaxy. We note that this is intended to investigate the direct effect of including near-IR data on mass-to-light measurement (see Figure~\ref{fig:m2l_filters}), as described above. This resolution is sufficient to identify individual clumps, but does not resolve them well (see bottom panels of Figure~\ref{fig:obser_image1} and ~\ref{fig:obser_image2}).  

We, therefore,  adopt the second set of images which is convolved to match the resolution of ACS/WFC FR647M images. This difference gives us significantly better resolution, which can be used to measure properties inside of clumps. There is a known systematic offset in the WCS positioning of the images from ACS and those from WFC3/UVIS, which we correct using standard point source matching methods. The FWHM in the images set to match FR647M is 0.05''. 

\subsection{Identification of Stellar Clumps}
\label{sec:clump_selection}
Historically, clumps have been identified in either emission line maps that trace star formation or in rest-frame blue wavelength data ($U$ or $B$ band). \cite{Lenkic_2021} show that significant color gradients exist inside of clumps, which implies that longer wavelength identifications of clumps will be systematically different. In this paper we aim to identify similar young structures as in $z>1$ systems \citep[e.g.,][]{Guo_2015,Jones_2010,Livermore2012}, and study the stellar mass associated with them. We, therefore, identify clumps using F336W ($U$-band) images. We use an unsharp masking technique similar to that developed by \cite{Fisher_2017} and not dissimilar from that used in CANDELS survey \cite{Guo_2015}. We opt for F336W as detection band rather than $NUV$-band (F225W) since the F336W image has a higher signal-to-noise ratio. Moreover, this is a similar rest frame wavelength to that used by \citep{Guo_2015}, $\sim$300~nm. 

\begin{figure*}
	\includegraphics[width=\textwidth]{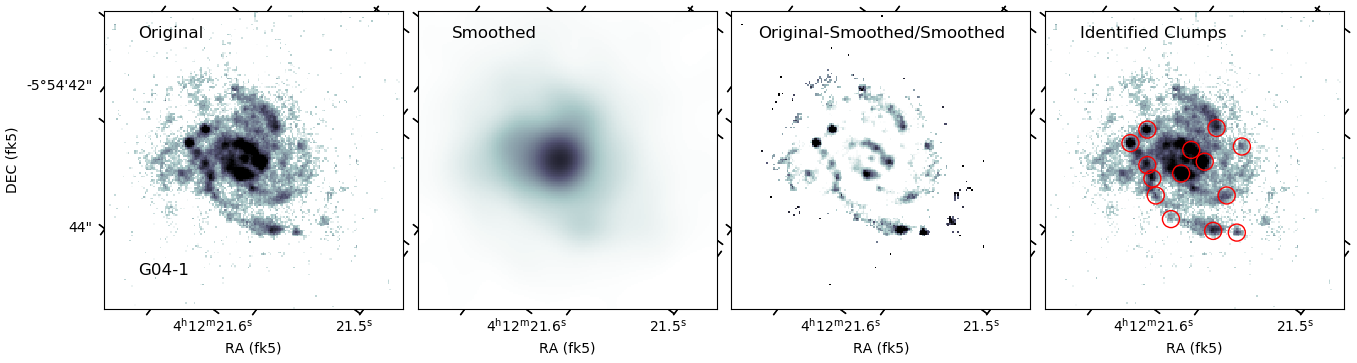}
    \caption{The above image illustrates various stages of clump identification in DYNAMO galaxies. Original image of galaxy G04-1 in F336W band, smoothed image that was created by convolving the original image by a gaussian with 8 pixel FWHM, unsharp masked image created by subtracting the smoothed image from the original image then dividing by the smoothed image. The last panel shows all identified clumps in this galaxy.}
    \label{fig:unsharp_masked}
\end{figure*}

\begin{figure*}
	\includegraphics[width=\textwidth]{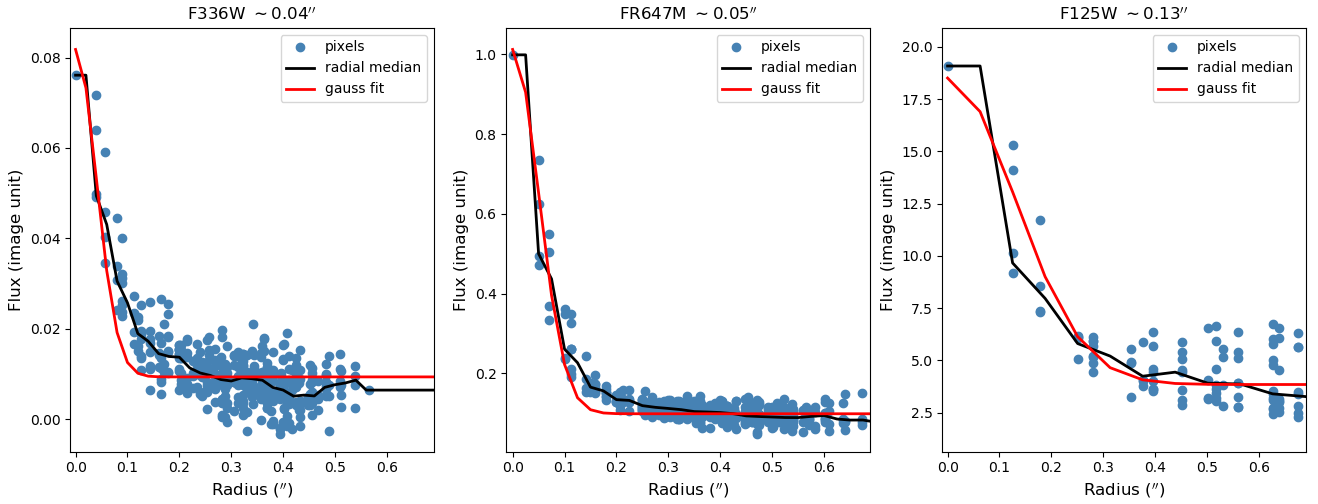}
    \caption{An example of light profile of a clump in F336W ($\sim$0.04''), FR647M ($\sim$0.05'') and F125W ($\sim$0.13'') from galaxy D13-5.  The red line indicates a 1-D Gaussian fit to the data, and the black line represents the median radial in radial bin fit to the data. This again shows how the F125W data resolved clumps poorly.}
    \label{fig:light_profile}
\end{figure*}

We refer the reader to Fig.~\ref{fig:unsharp_masked} where we use target G04-1 as an example to illustrate the clump selection method. First, the F336W images (first panel of Figure~\ref{fig:unsharp_masked}) were convolved with a Gaussian kernel with a FWHM that is 8 pixels.  Next, we subtract the convolved image (second panel of Figure~\ref{fig:unsharp_masked}) from the original and divide the difference image by the convolved image. This image  gives a detection image (third panel of Figure~\ref{fig:unsharp_masked}) that we use, in combination with the original image, to identify clumps. 

The clumps were identified as meeting the following four criteria: (1) at least 2$\times$ the background scatter within the region of a galaxy in the detection image; (2) at least a 5$\sigma$ peak above the galaxy disk light in the original F336W image; and (3) it must maintain (1) \& (2) over an area of at minimum 2$\times$2 pixels, a full resolution element; (4) clumps must be an independent source, that is the flux must decline in all directions from the local peak in emission. 

The cut values in we use in criteria (1) are chosen to be roughly consistent with a ``by-eye'' visual selection method. In the sample of six DYNAMO galaxies, we detected a total of 66 clumps. This corresponds to an average of 11 clumps per galaxy; however, the number of clumps per target ranged from 3 (G14-1) to 17 (D13-5). This is a similar number of clumps per galaxy as is observed in $z>1$ lensed observations \citep[e.g.,][]{Livermore2012,Cava_2018}. Our aim with DYNAMO work is to recover similar properties as are observed in higher-z systems, which again motivates this choice. Using a higher cut value will result in decreasing the number of clumps, and possibly biasing to higher masses. An alternate method of identifying clumps may simply be to calculate the $\Sigma_{SFR}$ and size in H$\alpha$ maps, it is beyond the scope of this work to investigate this particular systematic. 

Figure~\ref{fig:light_profile} shows an example of the light profile of a clump in the galaxy D13-5 in three filters from different resolutions. These show that the clumps we detected are well resolved in different resolutions and bands. 

\subsection{Measurement of Clump Sizes and Fluxes}
\label{sec:clump_measurment}
Once clump locations are identified in F336W band, we then use the F336W to measure the size of the clumps. We used F225W, F336W, F467M and FR647M to measure the flux of clumps. Our aim is to identify the young cluster and its extent in F336W, then determine its flux in all bands. We are therefore measuring the clump mass associated to young stellar population. This almost certainly introduces a systematic bias, as the size of clumps may vary as a function of stellar population. This age gradient inside of DYNAMO clumps has been studied by \cite{Lenkic_2021} on the same targets. Our clump sizes should be understood as the size of the young star forming region.

In order to measure the size of clumps we fit an elliptical Gaussian function to the 2D brightness distributions surrounding each peak in the F336W. The radius of each clump is then defined as the mean standard deviation of the major and minor axis of the 2-D Gaussian functions (i.e. $\approx FWHM/2.3553$). This size is the region in which a young stellar population is exceedingly bright compared to disks. We use the same measurement technique as was previously published in \cite{Fisher_2017}, see their section 3 for details. 

Clump fluxes are measured in the F225W, F336W, F467M and FR647M filters. The FR647M filter is a medium band ramp filter that was positioned to avoid the H$\alpha$ and [NII] emission lines. It is therefore a robust measurement of the $\sim$600~nm continuum flux. This provides a compromise between resolution and probing the older stellar populations. 
  
For each identified clump, we calculate the flux by integrating the light of each pixel within a defined aperture centered on each clump. However, multiple complexities occur when measuring clump fluxes. Overlapping flux from neighboring clumps (or the galaxy center) and the background light from the disk are both sources of systematic uncertainty in the clump flux. We construct a method that is similar to fitting Gaussians to clumps as a means to measure the flux, however it does not make {\em a priori} assumptions about the shape of clump surface brightness profiles. For each clump we define a square working region that is 1.2'' across, positioned at the center of each clump. We then measure the profile of the median flux as a function of radius for each clump in each filter. If a pixel flux is significantly brighter than the median, or if it is a systematic increase in flux with radius from the clump center, then that light is assumed to be representative of flux from the neighboring clump. These flux values are flagged and replaced by the median value at the same radial distance from the clump center. We find this gives very similar flux values as a Gaussian fit to each clump. We then integrate the "processed" clump flux within the clump radius (as defined in the F336W images).

In order to separate the light of clumps from the diffuse components of the galaxies, we subtract a local disk flux from each clump. To measure the light from the disk that is superimposed onto the clump, we fit a Gaussian plus constant function to each clump's radial profile in each band image after the masking has taken place, and use the constant as representative of the background disk light. The disk flux is then inward extrapolated to be that value over the same area covered by the clump. We note that for physical interpretation reasons, it is not clear if the disk should be removed from clump measurements or not. If disk light is subtracted, one is implicitly assuming that clumps are distinct objects from the background light, whereas if disk light is not subtracted, clumps are simply overdense regions within a disk. The latter argument is motivated by the observation that clump sizes are of order the same thickness as the disk. We therefore carry out both, and keep track the impact of local disk subtraction on our main results. We record all clumps both raw,  and processed and disk subtracted fluxes in each filter in Appendix \ref{app:clump_properties}, Table \ref{tab:flux_measurements}.

\begin{figure*}
    \includegraphics[width=\textwidth]{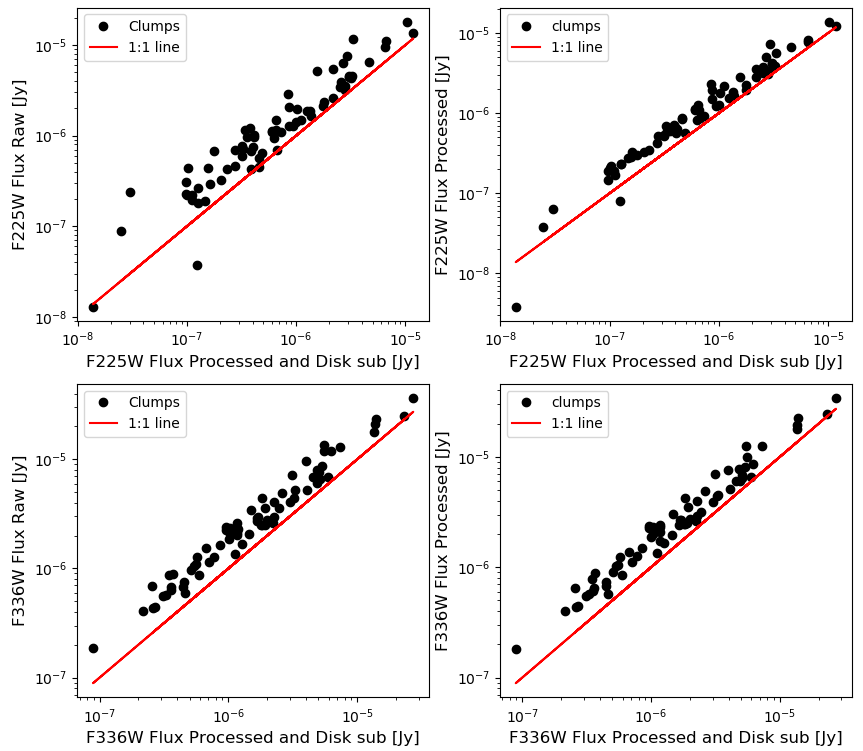}
     \centering
     \caption{Clumps fluxes comparison in F225W and F336W filters. Left panel: Comparison of clumps fluxes measured before (y-axis) and after processing and disk subtraction (x-axis). Right Panel: Comparison of clump fluxes after processing only (y-axis) and both processing and disk subtraction (x-axis).
     The red line indicates one-to-one relation between fluxes. The largest effect on clump flux is from disk subtraction, which reduces clump fluxes by $\sim$50\%.}
     \label{flux_comparsion1}
\end{figure*}

\begin{figure*}
    \includegraphics[width=\textwidth]{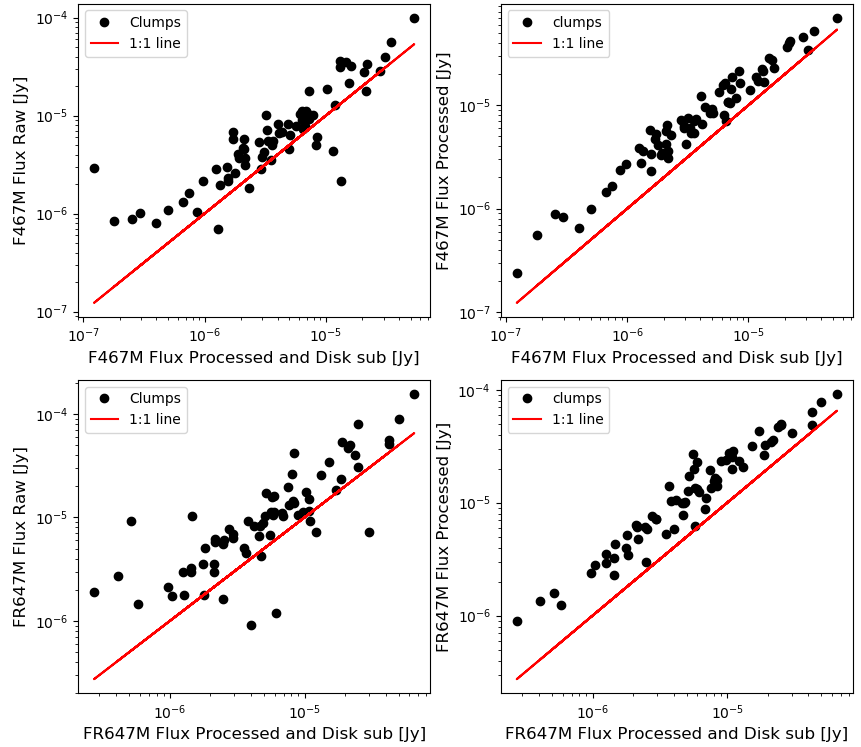}
     \centering
     \caption{The same as Figure~\ref{flux_comparsion1} but for clump fluxes in the F467M and FR647M filters, respectively}
     \label{flux_comparsion2}
\end{figure*}

In Figures~\ref{flux_comparsion1} and \ref{flux_comparsion2}, we assess the effect of "processing" and disk correction on the calculated flux of clumps in each filter for all our targets. We compare the flux of the clumps after the processing and disk subtraction to the raw clumps flux (before "processing" and disk subtraction) assuming the same size/diameter. We also compared the "processed" clumps flux but non-disk corrected with "processed" and disk corrected.  As is clear from the figure, the effects of both local disk subtraction and the processing procedure vary only mildly with clump flux, and do not vary significantly from galaxy to galaxy. Without disk subtraction, the processed and raw clump fluxes are consistent to 10\%. The impact of disk subtraction is more significant, reducing the flux by a median of roughly 50-60\% in the FR647M filter. The impact is more significant for the faintest clumps, whose fluxes are reduced by a factor of $\sim$3-4$\times$ , while the fluxes of the brightest clumps are reduced by a factor of 1-2$\times$ at most.

\subsection{Measuring Stellar Masses by SED Fitting}
\label{sec:mass_to_light_ratio}
We measured the stellar mass through fitting stellar population synthesis models to the observed SEDs. We used the CIGALE (Code Investigating GALaxy Emission) SED fitting software \citep{Boquien_2019} to derive physical properties of clumpy galaxies, such as stellar masses, age, and extinction. Here we note that, we performed the SED fitting on those two sets images described in (section \ref{sec:psf_convolution}): (1) We ran CIGALE in each resolution element in the galaxy using all available filters from our program from WFC3/UVIS (F225W, F336W, F467M), ACS/WFC (FR647M) and WFC3/IR (F125W) that were matched and resampled to the resolution of the F125W images. We remind our reader that this is only intended to assess the effect of including near-IR (F125W) data on our stellar mass measurement. (2) We ran CIGALE using clump fluxes measured from WFC3/UVIS (F225W, F336W, F467M), and ACS/WFC (FR647M), that were matched and resampled to the resolution of the FR647M images.  

\begin{figure*}
    \centering
	\subfloat{\includegraphics[width=\columnwidth]{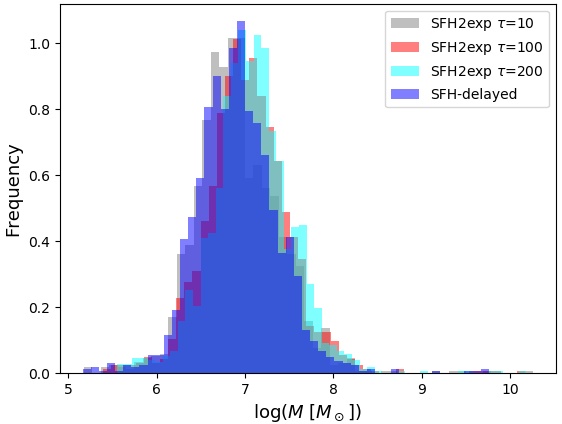}}
	\subfloat{\includegraphics[width=\columnwidth]{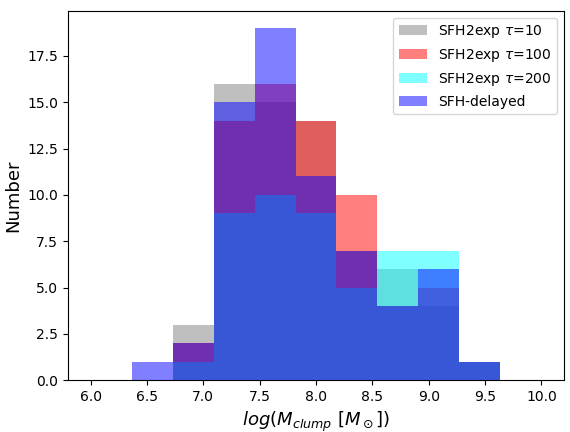}}
    \caption{Comparison of stellar mass distribution of every single pixel (left panel) and clump (right panel)in all DYNAMO galaxies using two different star formation history models. The blue one is with SFH-delayed model. The grey,red, and cyan is from the double exponentials model using three different fixed $\tau$ = $10, 100, 200$~Myr, respectively}
    \label{fig:allpixel_mass}
\end{figure*}

For the SED fitting, we adopted the stellar tracks of \cite{Bruzual_2003} at solar metallicity, which is close to the value measured in gas-phase metallicity with SDSS for our sample \citep{pettini2004}. We adopted the \cite{Chabrier2003} initial mass function. The \cite{Calzetti_2000} attenuation law is currently the most used for $z\sim 2$ galaxies. We implemented a modified Calzetti dust law and specific recipe described in \cite{Buat2011}. The authors analysed a sample of strong $UV$ emitting galaxies at $z\sim 2$, and found evidence for a steeper $UV$ slope than is reported \cite{Calzetti_2000}. Following \cite{Buat2011} and \cite{Faro2017}, we fixed the power-law slope to $-0.3$ and varied E($B-V$) from $0.05$ to $1.3$ for young populations. We allowed the age of the main stellar population to vary between $10-1000$~Myr , based on expectations from absorption line studies of DYNAMO galaxies \citep{Bassett_2014}.

In Figure~\ref{fig:allpixel_mass}, we consider two commonly used star formation history models available in the CIGALE SED fitting code. The first is a star formation history defined by double exponentials (\emph{sfh2exp}). In this model, a burst is superimposed on a decaying older stellar population \citep[see][for more in-depth discussion]{Boquien_2019}. In the \emph{sfh2exp} models, we used three different assumptions for the late burst component $\tau$=$10, 100, 200$~Myr. We also allowed the burst fraction to vary from 0.01-0.5 in each pixel. This has previously been used to measure stellar masses of high redshift galaxies \citep[e.g.,][]{Glazebrook_2004}. The second one is a delayed $\tau$ star formation history model, \emph{sfhdelayed}, which gives a nearly linear increase of star formation until the age (the time of onset of star formation) is equal to $\tau$, then decreases exponentially \citep{Boquien_2019}. This star formation history model is commonly used in the literature, including in analysis of clump masses  \citep[e.g.,][]{Dessauges_Zavadsky_2017,Cava_2018,Guo_2018}. In the \emph{sfhdelayed} model, we adopt $\tau$=10,30,50,100,200~Myr. We note that we used the same IMF and extinction properties for both star formation history models.

The histograms in Figure~\ref{fig:allpixel_mass} show the impact that these different star formation history models have on stellar masses. We find very similar stellar mass distributions for all runs. The median stellar mass of  all pixels derived from the star formation history model is log(M/$\mathrm{M}_\odot$)=$6.9$ with standard deviation $0.46$~dex. For the double exponential SFH model, we find that the median stellar mass of all pixels is log(M/$\mathrm{M}_\odot$) = $6.9\pm0.47$, $7.0\pm0.47$, $7.1 \pm 0.46$ for $\tau$ = $10, 100, 200$~Myr, respectively. These results essentially reiterate the result of \cite{Taylor2011}, in which they find that mass-to-light ratio is a robust quantity within SED modelling. 

We also made the same comparison for raw clumps stellar masses, and this is shown in the right panel of Figure~\ref{fig:allpixel_mass}. We find very similar clump stellar mass distributions for all runs. The median clump stellar mass with the SFH-delayed model is log(M/$\mathrm{M}_\odot$)=$7.75$ with standard deviation $0.61$~dex. For the SFH-double exponentials model, we find a median of clump stellar mass  log(M/$\mathrm{M}_\odot$) = $7.77\pm0.62$, $7.82\pm0.60$, $7.85 \pm 0.64$ for $\tau = 10, 100, 200$ ~Myr, respectively. Because we do not find a difference in mass distribution of each pixel and clumps, we opt for the simplest model, of the delayed $\tau$ models. 

Furthermore, we tested our method by running CIGALE for integrated galaxy light to measure the stellar mass, age, SFR, and dust extinction of the galaxies. From our integrated measurement, we found these physical properties to be very consistent with results from previous studies of DYNAMO galaxies, using SDSS magnitudes for stellar mass, H$\alpha$ for SFR, and Balmer line series for extinction \citep{Green_2014,Bassett_2014,Bassett2017,Fisher_2019}.

As previously mentioned, including near-IR pass-bands significantly increases the ability to derive robust stellar population properties from SED fitting methods \citep[e.g.,][]{Bell_2001,Zibetti_2009}. We note that measuring the near-IR light of individual clumps is unique to the DYNAMO sample, HST programs of $z>1$ galaxies cannot observe rest-frame near-IR. We therefore test the accuracy of the mass-to-light ratio determination using only optical pass-bands, which can also be a comparison to biases that would be present in HST surveys of $z\sim2$ galaxies. We carried out this test by simply re-running CIGALE without the WFC3/IR F125W filter flux and following the same procedure mentioned above. We then compared this to the mass-to-light ratio measurements using all bands (near-UV to near-IR). The mass-to-light ratio was determined simply by dividing the stellar mass by the light measured from FR647M band. 

\begin{figure}
	\includegraphics[width=\columnwidth]{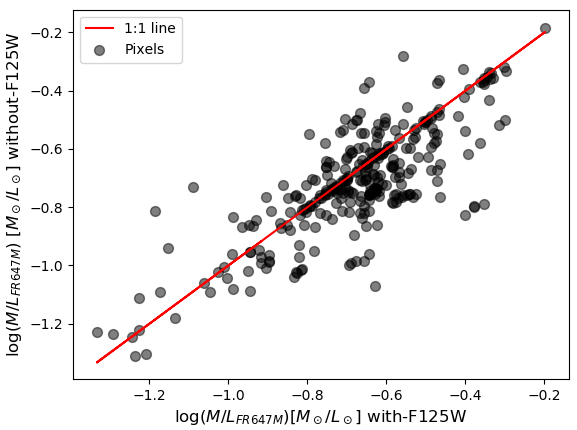}
    \caption{Comparison of the mass-to-light ratio in FR647M with and without near-IR (F125W) observations in every single pixel of the DYNAMO galaxy G04-1. The x-axis corresponds to the resulting mass-to-light ratio in the FR647M band when we use all five of our filters in our SED fit. The y-axis shows the mass-to-light ratio when we exclude the starlight from near-IR (F125W) observations in our SED fit. The mass-to-light ratios are nearly equal, which is shown as the red line representing the  one-to-one correlation between those mass-to-light ratio measurements.}
    \label{fig:m2l_filters}
\end{figure}

In Figure~\ref{fig:m2l_filters} we compare the FR647M mass-to-light ratio with and without the near-IR observations. The median $\log{M}/\mathrm{L_{FR647M}}$ is -0.65 with standard deviation of $0.18$ when we include near-IR starlight (F125W-filter). We find a median of $\log{M}/\mathrm{L_{FR647M}}=-0.71 \pm 0.19$ when we exclude the near-IR starlight. The difference between the median of $\log\mathrm{M}/\mathrm{L_{FR647M}}$ with and without F125W image is only 0.06 dex. This is visible in the plot, where there is only a slight tendency of $\log{M}/\mathrm{L_{FR647M}}$ to be lower when F125W filter is included. We note that for measuring clumps, this systematic bias is very small compared to the systematics introduced from the lower spatial resolution of F125W. 

Because we did not find a significant difference in mass-to-light ratio when we exclude the F125W band, we opt to determine the stellar mass of clumps using only the finer spatial scale images (WFC3/UVIS (F225W, F336W, F467M) and ACS/WFC (FR647M), which are matched and re-sampled to the FR647m resolution. We also remind the reader that the resolution of F125W reduces our ability to resolve individual clumps by a factor of $\sim 2\times$ (see also Figures~\ref{fig:obser_image1} and ~\ref{fig:obser_image2}). Therefore, we only run CIGALE to individual clumps using their raw fluxes, and processed and disk subtracted fluxes, to determine their stellar masses. We note that we used the same well tested CIGALE SED fit parameterization that is described in detail above.


\begin{figure}
	\includegraphics[width=\columnwidth]{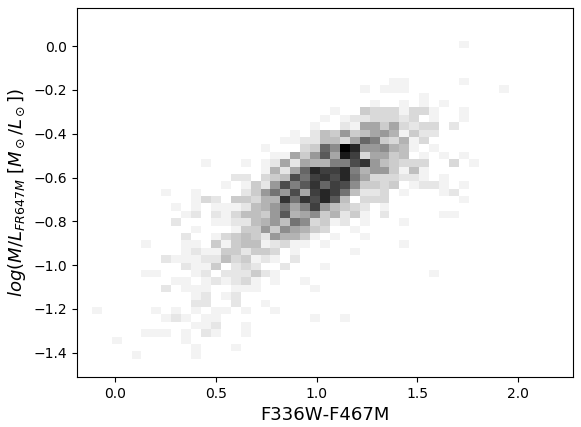}
    \caption{Correlation of the $F336W-F467M$ color and stellar mass-to-light ratio in FR647M band for each pixel in all DYNAMO galaxies.}
    \label{fig:color_m2l}
\end{figure}

In addition, we carried out a sanity check on how the mass-to-light ratio at a given band varies with the color in every single pixel in the galaxy, shown in Figure~\ref{fig:color_m2l}. In Figure~\ref{fig:color_m2l}, we show the correlation between all pixels rest-frame color $F336W-F467M$ ($U-B$), which is an indicator of age versus mass-to-light ratio in FR647M band. In general, we find that blue regions have lower $\mathrm{M}/\mathrm{L_{FR647M}}$ whereas the redder regions have higher $\mathrm{M}/\mathrm{L_{FR647M}}$ as expected for this type of galaxy. This result is quite similar to \citep{Zibetti_2009} found in the local universe.

\begin{figure}
	\includegraphics[width=\columnwidth]{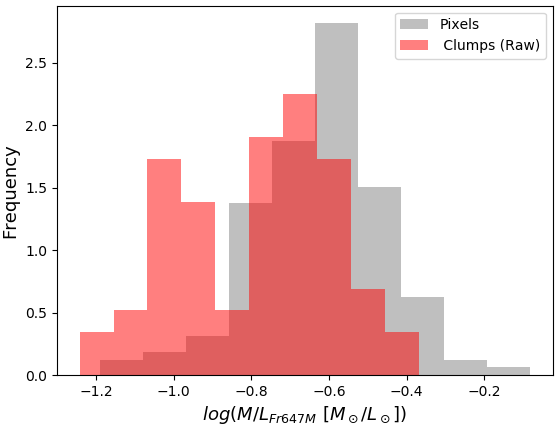}
    \caption{Distribution of mass-to-light ratio in FR647M of raw clump (red) in DYNAMO galaxies and every single pixel of DYNAMO galaxies(grey).}
    \label{fig:m2l_histo}
\end{figure}

In Figure~\ref{fig:m2l_histo}, we show the distribution of mass-to-light ratios in FR647M band for all raw clumps (red) and for all pixels in each galaxy disk (excluding clumps; grey). We determine the mass-to-light ratio of clumps by simply taking the stellar mass of the clumps from our SED fit and dividing by the light measured from the FR647M band. In order to make a one-to-one comparison of clump mass-to-light ratio to the disk mass-to-light ratio, we blur the images to the average clump size. We then re-run CIGALE using the new images following the same procedure as mentioned in Sect. \ref{sec:mass_to_light_ratio}. The pixels that are co-located with the clumps in the new image were excluded from the galaxy disk. The median mass-to-light ratio of the disk population is $-0.61$ with a standard deviation of $\pm 0.17$. 

Clumps are skewed towards lower mass-to-light ratios, with a median of $-0.75\pm0.20$. The distribution of clump $M$/$L_{FR647M}$ in DYNAMO galaxies is, though very broad. There appears to be a very low $M$/$L_{FR647M}\sim 0.1$ group of clumps and then a similar number of clumps that have comparable $M$/$L_{FR647M}$ to the disk light. The $M$/$L_{FR647M}$ peak is to some degree set by the fact that clumps are defined as peaks in blue wavelength light, and therefore younger populations. \cite{Lenkic_2021} found similarly that clumps are consistent with significantly lower ages compared to the disk, and the \cite{Lenkic_2021}  results would suggest that the low $M$/$L_{FR647M}$ values are due to clumps having younger stars, rather than less extinction. In general that clumps have low mass-to-light ratio than the disks. Clumps with $M$/$L_{FR647M}$ above -0.61 are $22\%$ of the entire population of clumps. 

\begin{figure}
	\includegraphics[width=\columnwidth]{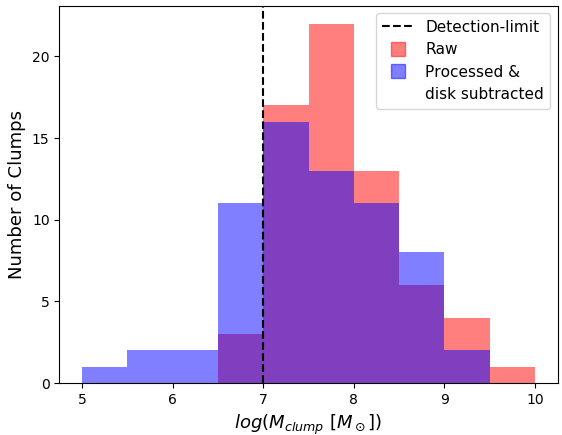}
    \caption{Stellar mass distribution of all clumps in our DYNAMO galaxies for raw clumps (red) and processed and disk subtracted clumps (blue). The vertical black line indicates our minimum mass detection limit in our targets, which is $\sim 10^{7}$ M$_{\odot}$. See Table \ref{tab:mass_measurements} in Appendix \ref{app:clump_properties} for raw clump stellar masses, and processed and disk subtracted clump stellar masses.}
    \label{fig:clumps_mass}
\end{figure}

\section{Result and Discussion}
\label{sec:clump_result}
\subsection{Stellar Mass of Clumps in DYNAMO Galaxies}

Figure~\ref{fig:clumps_mass} shows a distribution of the stellar masses of the raw clumps, and processed and disk subtracted clumps of the DYNAMO galaxies. We find that the clump mass in our sample ranges from $4.71 \times 10^{6}\ \mathrm{M}_\odot$ to $3.74 \times 10^{9}\  \mathrm{M}_\odot$ and $2.95 \times 10^{5}\  \mathrm{M}_\odot$ to $2.03 \times 10^{9} \mathrm{M}_\odot$ and that the average clump mass is $2.53 \times 10^{8}\mathrm{M}_\odot$ and $1.60 \times 10^{8} \mathrm{M}_\odot$ for raw clumps, and processed and disk subtracted clumps, respectively. 

The vertical solid line indicates our clump minimum mass detection limits, which is very near $\sim10^7$~M$_{\odot}$. To determine the minimum mass detection limit, we take 5 different background regions in each galaxy assuming an aperture size of $\sim 3\times3$ pixels, which is the average size of a clump. We measure the mass of each background region in the galaxy, and then calculate the average equivalent ``mass" of the 5 background regions in each target. We then take the average equivalent mass from each target as the minimum detection mass limit. We note that the actual mass detection limit will be different for each galaxy, which have different redshifts. This limit is intended to be a rough characteristic of the detection limit for the sample. From a total of 66 clumps, only two clumps and seven clumps are below our detection limit in raw and processed and disk subtracted clumps, respectively. We note that the peak of our distribution of masses is nearest to the detection limit, which suggests that in clumpy galaxies, clumps may be the high mass end of a continuum of masses that begins at lower masses.

Generally, we find similar mass distributions between the raw clumps, and processed and disk subtracted clumps. We obtain that on average, the processing and disk subtraction procedures reduce the clump masses by 40\%. We find a similar mass drop in all clumps. In addition, the impact of processing and disk correction on clump masses is similar across our targets. 

Overall, we find our distribution of clump masses favours low masses, as indicated both by the peak in the histogram (Figure~\ref{fig:clumps_mass}) and the slope of the distribution in the stellar mass function (Figure~\ref{fig:mass_function}). Theories in which clumps originate from ex-situ source, such as mergers and accretion, suggest that clumps would favour larger masses, or at least have an increase in high masses \citep[see discussion in][]{Huertas2020}. We do not find this in the DYNAMO clump masses. 

Results from simulation work show that stellar masses of clumps in turbulent, clumpy disk galaxies at $z\sim 2$ depend on the nature of the driver of stellar feedback assumed \citep[e.g.,][]{Mayer_2016,Mandelker_2016}. FIRE simulation results show that in high $\Sigma_{SFR}$ regions there is a strong radiation pressure, and this results in clumps that are short-lived $\leq$ 20~Myr \citep{Okl_2016}, see also similar results with the NIHAO simulation and the impact of so-called ``early stellar feedback" \citep{Buck_2017}. \cite{Mandelker_2016} discuss the impact of radiation pressure prescriptions on the distribution of clump masses. In effect, only clumps that are very massive are able to survive the strong radiative feedback models. This would result in a mass distribution of clumps that is skewed to high masses, and the vast majority of clumps are disrupted by internal feedback before they are able to migrate to the galaxy center. The DYNAMO sample tends to favour lower mass clumps. The most common mass of clumps in DYNAMO galaxies is 1-3$\times10^7$~M$_{\odot}$, depending whether the disk is subtracted or not.

Simulations in which stellar feedback is primarily driven by supernovae generate clumps that are long-lived with ages of $100-500$~Myr and with stellar masses greater than $10^{8} \mathrm{M}_\odot$ \citep[e.g.,][]{Bournaud_2014,Mandelker_2016}. More specifically, \cite{Mandelker_2016} finds intermediate mass clumps $10^{8-10}$ $\mathrm{M}_\odot$ systematically increase when they include early radiative feedback compared to simulation runs without radiation pressure. The clump stellar mass in DYNAMO galaxies falls in the range of $10^{6-9} \mathrm{M}_\odot$, which is consistent with these simulation results. Our result is in good agreement with their "only supernova feedback" recipe.

Observational studies at $z\sim 2$ such as \citet{Elmegreen_2008,Schreiber_2011,Guo_2012} obtained clumps masses of $10^{8-9} \mathrm{M}_\odot$. Recent work by \citet{Huertas2020} on the CANDELS survey argues that observational effects such as, resolution could lead to an overestimation of clumps masses by a factor of 10, which is similar to previous estimations \citep{Dessauges_Zavadsky_2017,Fisher_2017,Cava_2018}. This would be consistent with the difference in masses of a typical DYNAMO clump and those of $z\approx 1-2$ galaxies.

\subsection{Clumps Stellar Mass Function}
\begin{figure}
	\includegraphics[width=\columnwidth]{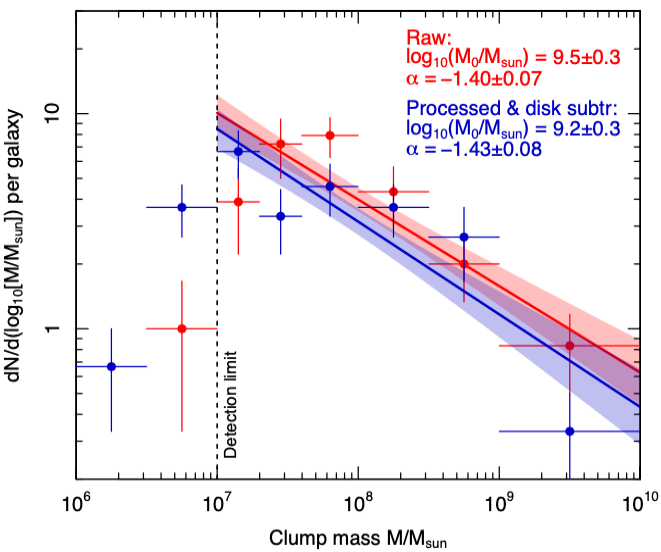}
    \caption{Stellar mass function of all clumps in DYNAMO galaxies. The circle indicates the resulting stellar mass function distribution of raw and processed and disk subtracted clumps, respectively. The solid and dash line indicate a single power-law fit to the mass. The vertical black line in both panels indicates our minimum mass detection limit in our targets, which is $\sim 10^{7}$ solar mass.}
    \label{fig:mass_function}
\end{figure}
The rate of change in the distribution of clump masses has been studied by other authors, and is interpreted to imply clumps origin. For example, in the simplest scenario {\em in situ} clump origins are expected to have a more steeply declining rate of change than {\em ex situ} clump origins, in which some estimate an up turn in the rate of change at higher masses.

We can characterize the distribution of clump stellar masses $M$ by a mass function (MF), $\Phi(M)$. This function is defined such that the expected number of clumps per galaxy in our sample, in a small stellar mass range $[M,M$+$\delta M]$, is
\begin{eqnarray}
    \delta N = \Phi(M)\delta M.
\end{eqnarray}

It follows that the expected number of clumps per galaxy in the range $[x,x+\delta x]$ of logarithmic mass $x=\log_{10}(M/\mathrm{M}_\odot)$ is
\begin{eqnarray}
\delta N = \phi(x)\delta x,
\end{eqnarray}

with $\phi(x)=\ln(10)M\Phi(M)$. In consideration of the small galaxy sample, we limit our analysis to fitting a power law MF,
\begin{eqnarray}
	\Phi(M) = \frac{1}{M_0}\left(\frac{M}{M_0}\right)^\alpha,
\end{eqnarray}
with only two free parameters, the characteristic mass scale $M_0$ and the exponent $\alpha<0$. In logarithmic terms, this model reads

\begin{eqnarray}
    \phi(x) = \ln(10)\left(\frac{M}{M_0}\right)^{\alpha+1}
\end{eqnarray}

Fitting such a MF to a discrete set of observed masses $M_i$ (with $i$ being the clump index) is a tricky problem, especially if the sample is relatively small. Perhaps the most intuitive method, still often used in fitting galaxy mass functions, is to bin the data into regular mass or log-mass intervals and then fit the MF to the binned data using a standard fitting technique such as $\chi^2$-minimisation. The downside of this approach is that it depends on the arbitrary choice of bins and often behaves badly if the bins are gradually reduced to infinitesimals. Only by carefully expressing the likelihood function using predictive Poisson statistics see \citep[e.g.,][]{Cash_1979} in infinitesimal bins, can we make the bins disappear correctly. The exact likelihood function of the clump MF then becomes

\begin{eqnarray}
\label{eq:LL}
 	\ln L = \sum_i \ln\left[N\phi(x_i)\right]-N\int\phi(x)dx,
\end{eqnarray}
where $i$ goes over all the clumps and $N=6$ is the total number of galaxies. This number appears in the likelihood, because we have defined the MF in such a way that it returns the distribution of clump masses \emph{per galaxy} in the sample. For a full derivation of eq.~(\ref{eq:LL}), please refer to equations (3)--(10) in \citet{Obreschkow_2018} (note that their effective volume $V(x)$ is analogous to the number $N$ in the present case). This reference also explains how the likelihood can be generalised to account for mass measurement uncertainties. We here neglect such uncertainties, but note that identical and normally distributed measurement errors would only affect the power law normalisation $M_0$, not the index $\alpha$.

We determine the free parameters $M_0$ and $\alpha$ by maximising eq.~(\ref{eq:LL}), while evaluating the integral between $x=7$ (the detection limit) and $x=10.5$ (the approximate value of the galaxy stellar masses). The fits are obtained via the \textit{dftools} package published by \cite{Obreschkow_2018} for the $R$ statistical language. This approach readily returns the maximum likelihood solution, as well as parameter uncertainties and Bayesian evidence estimates.

The best-fitting power law solutions with their 1-$\sigma$ uncertainty ranges, obtained by propagating the covariance matrix of $M_0$ and $\alpha$, are shown in Figure~\ref{fig:mass_function}. We show the stellar mass function of the raw (red) and processed and disk subtracted clumps (blue) in DYNAMO galaxies. The black vertical line indicates our clumps minimum masses detection limit in DYNAMO galaxies. In our power-law fitting, we exclude clumps that have masses below our minimum masses detection limit to avoid incompleteness issue. We find a power-law slope of $-1.40\pm 0.07$ and $-1.43\pm 0.08$ for raw and processed and disk subtracted clumps. 


\cite{Dessauges_2018} calculated the stellar mass function of clumps in a sample of lensed galaxies at $z\sim1-3$. They find slopes more similar to $-1.7$. This is within uncertainties from the mass function power-law in local spirals, $\alpha\approx -2\pm0.3$ \citep{Adamo_2013}. \cite{Dessauges_2018} argue that small, inhomogenous samples, similar to what we study with DYNAMO, can lead to shallower slopes than the true underlying populations of star clusters in galaxies.
\cite{Dessauges_2018} used a sample of 27 galaxies with 194 clumps, which is $\sim3\times$ larger than ours. Our sample, has the feature of being more consistent in selection of galaxy mass, and all are well studied as rotating, marginally stable disks \citep{Fisher_2017b}. Recently, \cite{Huertas2020} finds a power-law of $\alpha\approx -1.55 \pm 0.34$ for data from the VELA simulations. We do not know if the differences from DYNAMO sample is significant or due to the low number statistics of this difficult to determine quantity.

 
\begin{figure}
	\includegraphics[width=\columnwidth]{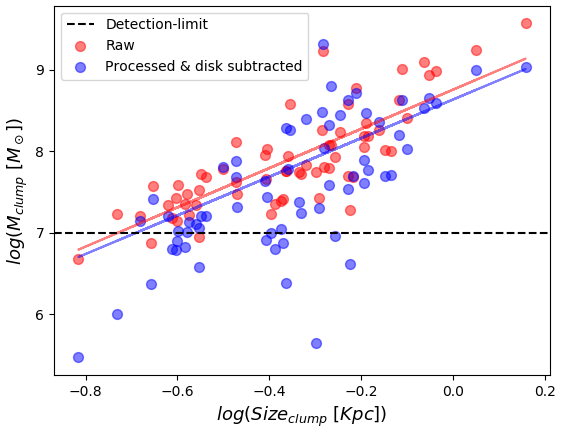}
    \caption{The clump stellar mass as a function of size of raw clumps (red) and processed and disk subtraction (blue) in DYNAMO galaxies. The red and blue line shows single power law fit to the correlation for the raw and processed and disk subtraction, respectively. The horizontal dash line indicates our minimum mass detection limit in DYNAMO galaxies, which is $\sim 10^{7}$ solar mass.}
    \label{fig:size_mass}
\end{figure}

\begin{figure*}
    \centering
	\includegraphics[width=\textwidth]{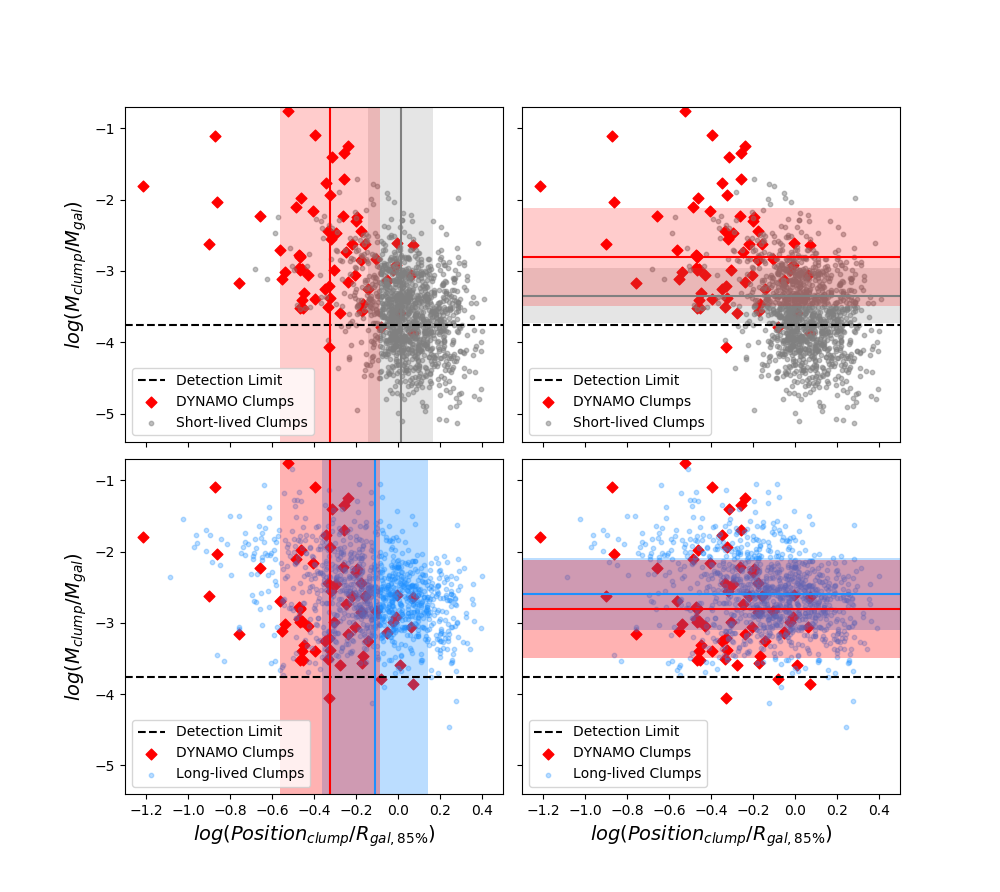}
    \caption{The gradient of clumps stellar masses with clumps position. The x-axis is clumps distance from the center of the host galaxy normalized by a radius that contains 85\% starlight of the host galaxy. The y-axis is clumps stellar mass normalized by stellar masses of the host galaxies. In all panels DYNAMO clumps are indicated in red symbols. In the top row we show comparison of DYNAMO clumps to short-lived (grey), and in bottom row we compare to long-lived clumps (blue) from simulation by \citet{Mandelker_2016}. The colored vertical (left) and horizontal line (right) indicates median of position (left) and stellar mass (right). The shaded area in all panels represents $\sim 68\%$ all respective samples position(left) and stellar mass(right). The shaded red and grey area encloses $\sim 68\%$ of DYNAMO and short-lived position of clumps, respectively. The vertical red and grey line indicates the median of  DYNAMO and simulated short-lived clumps position, respectively. On the top right panel, we compared the DYNAMO clumps to simulated short-lived clumps in terms of stellar mass. The shaded  area encloses $\sim 68\%$ of DYNAMO and short-lived and long-lived stellar mass of clumps. The dash horizontal line in all panels indicates DYNAMO galaxy minimum mass detection limit.}
    \label{fig:radius_mass}
\end{figure*}

\subsection{Mass-Size Relation for Clumps}
The size-luminosity and size-mass relationship for clumps has been studied by a number of authors \citep[e.g.,][]{Livermore2012,Wisnioski_2012,Fisher_2017, Cava_2018,Cosens_2018,Messa_2019}. In general both H$\alpha$ flux and stellar mass have been shown to be at least roughly consistent with constant surface brightness or constant surface density. \cite{Fisher_2017} shows that the zero point of this correlation is offset and varies with galaxy SFR. 

In Figure~\ref{fig:size_mass}, we show mass and size of all clumps in DYNAMO galaxies for both raw (red) and processed and disk subtracted (blue) clumps. The horizontal dash line indicates our minimum clump mass detection limit in DYNAMO galaxies. A power-law fit into the data results in slope $\sim 2.4 \pm 0.22$ and $\sim2.3 \pm 0.30$ for raw and processed and  disk subtracted respectively. These slopes are calculated for clumps masses above our minimum mass detection limits.  We do not find a significant difference in this correlation due to effects of disk subtraction, aside from the known offset in masses. In general our work is consistent with that of previous work \citep[e.g.,][]{Cava_2018}. We note that for DYNAMO clumps we have now measured roughly the same power slope in L$_{H\alpha}-r_{clump}$ \citep{Fisher_2017} and M$_{star}$-$r_{clump}$, implying this relationship is similar, between the conversion of star formation into mass. On some level this is not surprising given the young ages of clumps. We also add the caveat that clump sizes are measured in F336W, which is dominated by young stars. \cite{Lenkic_2021} investigate the color gradients within clumps. They show that the sizes of clumps increase for redder wavelengths. Our aim in this paper is to characterise the stellar mass of the young star-forming clump, and this motivates the choice of F336W. 

\subsection{Galactocentric Distance Gradient in Clump Stellar Mass}

In Figure~\ref{fig:radius_mass}, we show the measured stellar masses of all DYNAMO clumps normalized by the host galaxy stellar mass as a function of clump position within the galaxy normalized to the radius that contains $85\%$ of the disk flux in the F125W starlight images. 

The figure shows a weak trend of clump mass with galaxy position. Where towards the galaxy center clump masses are larger ($M_{clump}/M_{gal}\sim1\%$) and smaller clumps are more common at large radius. However, we note that the Pearson’s correlation coefficient indicates a weak correlation at best (r=-0.30), and massive clumps ($M_{clump}/M_{gal}>1\%$) are observed to radii as large as $\sim60$\% of the galaxy light. Moreover, any trend in which more massive clumps are preferentially found in the galaxy center could be a selection effect, as lower mass clumps could blend with the disk in the galaxy center. A more informative comparison may come by comparing statistical averages to simulation results.

We compared our observations with the results from simulation by \citet{Mandelker_2016} in the same figure. The simulation does not account for disk subtraction, nor anything similar to our processing algorithm. We, therefore, only consider the raw mass of clumps in this comparison. We note that this correlation remains true in the processed and disk subtracted DYNAMO clumps, taking into account those systematic differences described above. We also point out the many systematic uncertainties in comparing galaxy substructure with simulation data, as there are very significant differences in how the physical properties are determined. One source of systematic uncertainty is in the normalizing galaxy sizes. The $R_{gal,85\%}$ for DYNAMO galaxies were estimated using the F125W star-light, whereas, in the simulations \cite{Mandelker_2016} used the ``cold mass", which will include stars and gas. We find typical $R_{gal,85\%}\sim 4-6$~kpc, which is similar to the disk sizes reported in these simulation \citep{Mandelker_2014}.

In all panels, DYNAMO clumps are represented by red symbols. In the top row we compare DYNAMO clumps to short-lived clumps from the simulation, and in the bottom row we compare to long-lived clumps. \cite{Mandelker_2016} defines the boundary between long-lived and short-lived clumps as $20\times$ the free-fall time, which would typically correspond to $\sim100-200$~Myr. The left and right panels show shaded regions and colored lines corresponding to the medians and $\sim 68\%$ of the respective samples in position (left) and mass (right).


First, we compare DYNAMO clumps to the short-lived clumps in the \cite{Mandelker_2016} simulation. DYNAMO clumps are neither at a similar location within the disk, nor do they have a similar range of masses to those of short-lived clumps.  We obtain a median of positions, log$(position_{clump}/R_{gal,85\%})$, for DYNAMO clumps of $-0.32$ with a standard deviation of $0.24$. We find $0.01\pm 0.15$ for the simulated short-lived clumps. There is almost no overlap between the majority of DYNAMO clumps and the majority of short-lived clumps, and short lived clumps are typically located at a radius that is a factor of $2\times$ larger than the location of the DYNAMO clumps. Similarly the median stellar mass of DYNAMO clumps is  log$(M_{clump}/M_{gal})$ is $-2.8\pm 0.68$, which is significantly higher than the median of  $-3.3 \pm 0.4$ for simulated short-lived clumps. Together the DYNAMO clumps are both more massive and located in more centrally concentrated regions that the short-lived clumps in the VELA simulation.

We now consider the populations of simulated long-lived clumps, shown in the bottom row of Figure~\ref{fig:radius_mass}. Long-lived clumps are a much better match the position and mass of clumps in DYNAMO galaxies. We obtain a median of positions log$(position_{clump}/R_{gal,85\%})$ for simulated long- lived clumps of $-0.11 \pm 0.25$. There is overlap between the majority of DYNAMO clumps and the majority of long-lived clumps, and the difference between the median position of DYNAMO clumps is $\sim 0.21$ dex, which is within the scatter of both. The median stellar mass log$(M_{clump}/M_{gal})$ is $-2.56 \pm 0.5$ simulated long-lived clumps. The difference between the median stellar mass of DYNAMO and long-lived clumps is roughly $\sim 0.2$ dex.  


The majority of simulated short-lived clumps are faint and reside in the outskirts of the disk compared to simulated long-lived clumps. This may bias our comparison, due to the sensitivity limits of the HST data. Low flux clumps in the outskirts of DYNAMO disks could go undetected. We, therefore, apply a positional cut at log$(position_{clump}/R_{gal,85\%})$ < $0$ for both in comparison with position and stellar mass.  We note that all the above values for both the simulation and observations are determined only for the clumps that have stellar mass greater than our minimum mass detection limits (indicated in a black horizontal dash line in Figure~\ref{fig:radius_mass}).

We obtain a median of position log$(position_{clump}/R_{gal,85\%})$ is $-0.33\pm 0.23$ for DYNAMO and $-0.07\pm 0.11$  for simulated short-lived clumps. We find median of stellar mass log$(M_{clump}/M_{gal})$ is $-2.79 \pm 0.69$ and $-3.25 \pm 0.43$ for DYNAMO clumps and simulated short-lived clumps, respectively. The difference between the median positions DYNAMO and short-lived is $\sim 0.26$ dex and stellar mass is $\sim 0.45$, which is very close to median difference before we apply the positional cut. There is no significant change in the distribution of masses. There may indeed be a population of low mass short-lived star clusters in the outer disks of DYNAMO galaxies, however the large clumps that are detected in a way that is intended to be similar to clumps in $z>1$ are not similar in properties to the short-lived clumps in VELA simulations.

Recently, \cite{Lenkic_2021} studied the internal gradients of optical colors in DYNAMO galaxies using the same HST observations as we present here. They found results consistent with a complex substructure of ages, in which the center of the clump is young and the outer part is old. They interpret these observations as indicating that clumps are long-lived structures, which contain an internal centrally concentrated star formation event. This seems similar to the description of clumps in the simulations of \cite{Bournaud_2014} in which clumps are long-lived and the ages based on optical star-light are continuously rejuvenated by supplies of fresh new gas. There is therefore a building of evidence that clumps in DYNAMO galaxies are more consistent with properties of long-lived clumps. 

From this comparison, we conclude that the properties of DYNAMO clumps are inconsistent with the properties of simulated short-lived clumps in the VELA simulation \citep{Mandelker_2016}. We find the majority of DYNAMO clumps are closer to the center than simulated short-lived clumps and also more massive than simulated short-lived clumps. Generally, DYNAMO clumps properties are more similar to simulated long-lived than short-lived clumps, in position and stellar mass.

\subsection{Specific Star Formation Rate of Clumps in DYNAMO Galaxies}

DYNAMO galaxies are, by selection, high specific star formation rate systems. They are typically more consistent with values found in $z\approx1.5$ galaxies, rather than local spirals. \cite{White2021} show that this remains true for the resolved relationship of $\Sigma_{SFR}-\Sigma_{star}$ for one DYNAMO galaxy observed with Keck adaptive optics. They find $\Sigma_{SFR}$ that are a factor of $\sim2\times$ larger than the averages for $z\approx1$ galaxies from the CANDELS survey \citep{Wuyts2013}. We intend to complete a full analysis of clump specific star formation rates on our  HST sample in a future work (Ambachew {\em in prep}), but for the purposes of discussion we will consider averages here.

We estimate the average sSFR of clumps by combining SFR measured in $H\alpha$ observation from \cite{Fisher_2017} and stellar masses of clumps from this work. We find that in DYNAMO galaxies the average sSFR of clumps is  $4.1~Gyr^{-1}$, with an overall range of values from $\sim$0.2-7.5~Gyr$^{-1}$. This is within the range of values observed in $z\sim1-2$ galaxies \citep{Wuyts2013}. \cite{Mandelker_2014} studies the properties of clumps in $z\sim 2$ disk galaxies using Adaptive Mesh Refinement (AMR) cosmological simulation. They find the \textit{in-situ} clumps to have higher sSFR that ranges between $1-10~Gyr^{-1}$. Our preliminary estimation of sSFR clumps in DYNAMO galaxies is consistent with this simulation result. We will investigate this similarity in detail in our next work.

\subsection{Implications for Clump Evolution}
\cite{Dekel2013} outline a picture in which the evolution of clumps in galaxies is largely driven by the accretion of surrounding material in the disk and the subsequent outflows due to star formation. Simulations by \cite{Bournaud_2014} describe a similar picture in which clumps survive due to constant refresh of gas, in-spite of the mass loss due to strong outflows associated to high SFR surface density regions. While DYNAMO galaxies are very similar, we have nonetheless determined that the mass at DYNAMO resolution is lower than early estimates of the clump masses. This may have an impact on clump evolution.

Following \cite{Dekel2013}, we can check the consistency of DYNAMO galaxies with this framework, in light of the results here. In order for a clump to survive in a disk, the mass accretion rate ($\dot{M}_{acc}$) must be greater than the outflow mass rate ($\dot{M}_{out}$) over the lifetime of the clump. We can first estimate $\dot{M}_{out}$ from \cite{Dekel2013} and compare to outflow rates in similar galaxies. 

They state that $\dot{M}_{out}\approx \eta \epsilon_{ff} f_{gas} M_{clump}/t_{ff}$. While the fiducial assumption for star formation efficiency per free-fall time is low, $\sim$0.01, \cite{Fisher2022} recently found a higher value of $\sim$0.1 in a clumpy, turbulent galaxy similar to DYNAMO galaxies, and is similar to what is observed in nearby super-star clusters. We adopt this value for the efficiency per free-fall time in our mass outflow rate estimates. We adopt $\epsilon_{ff}\approx 0.1$ and similarly from \cite{Fisher2022} that $t_{ff}\approx 3-10$~Myr. Typical mass-loading factors, $\eta$, vary in clumpy galaxies. Ionised gas observations suggest $\eta\approx0.5$ \citep{Davies2019,Reichardt2022}, adjusting for molecular gas we adopt $\eta\approx1$. We can then take the typical DYNAMO gas fraction as f$_{gas}\approx 0.2$ and the average clump mass from this paper of $3\times10^8$~M$_{\odot}$. Taken together this gives a mass outflow rate for clumps of $\dot{M}_{out}\approx0.5-1$~M$_{\odot}$~yr$^{-1}$. This is similar to observations of mass-outflow rates in clumpy, star forming disk galaxies \cite[e.g.][]{Reichardt2022}. 

The mass accretion rate of clumps in DYNAMO galaxies is challenging to measure. We will follow \cite{Dekel2013} and determine if this is below the estimated outflow rate. They derive $\dot{M}_{out}\approx \alpha/2 (t_{ff}/t_d) M_{clump}/t_{ff}$. In this case both $\alpha$ and $t_{ff}/t_d$ are taken as constants of $\sim1/3$. For the same assumption on clump mass and $t_{ff}$ we derive a comparable $\dot{M}_{acc}\approx 1-2$~M$_{\odot}$~yr$^{-1}$. This is very similar to the outflow rates but higher and, if true, implies that many clumps would be consistently long-lived as they slowly accrete more mass than they expel over their lifetime.  

Multiphase outflow measurements of clumps in galaxies like DYNAMO (and $z\approx1.5$ disks) are a direly needed observation. More work in this area, perhaps with JWST and ALMA would be critical to study the evolution of clumps in this important phase of galaxy evolution.

\section{Summary}
We have studied the stellar masses of clumps in gas-rich, turbulent disk galaxies from the DYNAMO sample, which are similar in properties to $z\sim 1.5$ galaxies. DYNAMO galaxy observations allow us to study clumps with reduced uncertainty due to  finer spatial resolution and measuring the star light at longer wavelengths where the mass-to-light ratio is more robust against extinction and age effects. We used a sample of six DYNAMO galaxies observed with HST and identified 66 clumps in F336W (young stars). 

We find the stellar mass of DYNAMO clumps ranges from $0.04- 37.4 \times 10^{8} \mathrm{M}_\odot$ (alternatively $0.002 - 20.3 \times 10^{8} \mathrm{M}_\odot$ for disk subtracted clumps). This is consistent with the finding of clumps stellar mass in observation of lensed galaxies \citep{Cava_2018}. This is also consistent with VELA simulation mass threshold when they use only supernova as a feedback recipe \citep{Mandelker_2016}. We measured the power-law slope of the stellar mass function of the entire sample of clumps to be $\alpha -1.40\pm 0.07$, with little dependence on disk subtraction. This declining power-law slope is consistent with simulation work by \cite{Huertas2020}. We also observed a clear trend with clumps stellar mass increasing with the size of the clump. This is consistent with the findings of \citep{Cava_2018} in gravitational lensed galaxies.

We compare our observations with results from the VELA simulation \citep{Mandelker_2016}. Specifically we compare clump stellar mass to the position in the galaxy. \cite{Mandelker_2016} revealed  that the lifetime of the clump is connected to its position, where clumps in the central $\sim$50\% of the galaxy are almost always "long-lived" clumps, and short-lived clumps are restricted to a large radius. Moreover, the stellar masses of long-lived clumps are systematically higher than those of short-lived clumps by a factor of a few. We find the masses and galactocentric positions of DYNAMO clumps are inconsistent with simulated short lived clumps. However, clumps in DYNAMO galaxies are more similar to simulated long-lived clumps than simulated short-lived clumps. 

Observations of clumps in rest-frame near-IR light will soon be possible with JWST, at least at $z\sim1$. Our results, therefore can be tested in upcoming GTO and ERS programs. We note, however, that even with JWST the FWHM of restframe J-band light is of order 0.6-0.8~kpc at $z\sim1.5$, which is larger than the typical size of clumps in either DYNAMO galaxies or lensed galaxies. It will, therefore, still remain challenging to isolate individual clumps. More work is still needed to determine how the biases in resolution may impact upcoming results from JWST.

\section*{ACKNOWLEDGEMENTS}

We are grateful to Nir Mandelkar for making simulation results available to us. DBF is thankful to Sarah Busch for technical help. DBF acknowledges support from Australian Research Council (ARC)  Future Fellowship FT170100376 and ARC Discovery Program grant DP130101460. ADB acknowledges partial support form NSF-AST2108140.
Parts of this research were supported by the Australian Research Council Centre of Excellence for All Sky Astrophysics in 3 Dimensions (ASTRO 3D), through project number CE170100013. KG  is a recipient of an Australian Research Council Laureate Fellowship (
FL180100060) funded by the Australian Government DO is a recipient of an Australian Research Council Future Fellowship (FT190100083) funded by the Australian Government. I.D acknowledges the support of the Canada Research Chair Program and the Natural Sciences and Engineering Research Council of Canada (NSERC, funding reference number RGPIN-2018-05425).

\section*{DATA AVAILABILITY}

The DYNAMO HST data used in this paper are part of the Cycle 25 program 15069 and Cycle 20 program 12977, and are publicity available on the HST archive (\url{https://archive.stsci.edu/hst/search.php}).



\bibliographystyle{mnras}
\bibliography{references} 



\newpage
\appendix

\section{Clump Properties} \label{app:clump_properties}

\begin{center}
\onecolumn 
\begin{longtable}{lccccccccccc}
\caption{Fluxes of clumps in DYNAMO galaxies both raw and processed and disk subtracted in all filters}\\
\toprule
\label{tab:flux_measurements}
ID &RA&Dec&${F}_{f225w}$&${F}_{f225w}$ &${F}_{f336w}$&${F}_{f336w}$&${F}_{f467m}$&${F}_{f467m}$&${F}_{fr647m}$&${F}_{fr647m}$ 
\\&& & (Raw)&(Disk sub)& (Raw)&(Disk sub)& (Raw)&(Disk sub)& (Raw)&(Disk sub)&\\
\\ D13-5 & & & [$\mu Jy$]&[$\mu Jy$]&[$\mu Jy$]&[$\mu Jy$]&[$\mu Jy$]&[$\mu Jy$]&[$\mu Jy$]&[$\mu Jy$]\\
\midrule
1 & 13:30:07.120 &  +00:31:54.07 &  3.56 &      2.82 &   6.85 &      5.92 &   8.97 &      6.32 &   11.12 &       6.84 \\
2 & 13:30:07.173 &  +00:31:53.59 &  1.01 &      0.41 &   2.09 &      1.04 &   4.64 &      2.07 &    7.23 &      30.29 \\
3 & 13:30:07.079 &  +00:31:52.35 &  1.86 &      1.35 &   4.40 &      3.24 &   8.18 &      4.86 &   11.35 &       5.98 \\
4 & 13:30:07.061 &  +00:31:52.11 &  3.43 &      2.51 &   5.20 &      3.29 &   7.80 &      6.73 &    8.28 &       4.69 \\
5 & 13:30:07.046 &  +00:31:52.12 & 11.16 &      6.61 &  20.90 &     13.68 &  34.02 &     21.88 &   41.72 &       8.37 \\
6 & 13:30:07.051 &  +00:31:52.22 &  9.57 &      6.53 &  17.87 &     13.51 &  28.04 &     20.77 &   34.22 &      15.23 \\
7 & 13:30:07.013 &  +00:31:54.29 &  1.11 &      0.73 &   2.79 &      1.99 &   5.51 &      3.68 &    9.15 &       3.81 \\
8 & 13:30:06.988 &  +00:31:54.32 &  1.22 &      0.38 &   4.04 &      2.28 &  10.06 &      3.20 &   19.52 &       7.48 \\
9 & 13:30:06.956 &  +00:31:53.97 &  1.15 &      0.34 &   3.43 &      1.49 &   9.32 &      7.32 &   17.03 &       5.18 \\
10&  13:30:06.990 & +00:31:51.91 &  1.50 &      0.65 &   3.59 &      1.92 &   7.80 &      5.69 &   10.23 &       1.47 \\
11&  13:30:06.979 &  +00:31:52.31 & 1.96 &      1.02 &   4.96 &      2.60 &  11.26 &      6.43 &   18.44 &      17.10 \\
12&  13:30:06.927 &  +00:31:52.84 & 0.22 &      0.10 &   0.70 &      0.25 &   1.97 &      1.34 &    2.70 &       0.41 \\
13&  13:30:06.905 &  +00:31:53.38 & 0.44 &      0.10 &   1.87 &      1.01 &   5.33 &      2.79 &   10.45 &       8.97 \\
14&  13:30:06.938 &  +00:31:50.59 & 0.19 &      0.15 &   0.87 &      0.59 &   3.13 &      2.15 &    5.57 &       2.50 \\
15&  13:30:06.878 &  +00:31:50.96 & 0.47 &      0.27 &   1.16 &      0.71 &   2.18 &      0.97 &    3.26 &       1.44 \\
16&  13:30:07.199 &  +00:31:54.79 & 1.47 &      1.11 &   2.42 &      0.97 &   6.87 &      1.72 &    9.24 &       0.52 \\
17&  13:30:07.185 &  +00:31:56.12 & 0.20 &      0.11 &   0.45 &      0.27 &   1.10 &      0.50 &    1.47 &       0.58 \\
\bottomrule
    \toprule
     D15-3 \\
  	\hline
1 &  15:34:35.295 & -00:28:45.24 &   1.86 &      1.25 &   4.06 &      3.01 &   6.63 &      4.10 &   10.65 &       5.81 \\
2 &  15:34:35.308 & -00:28:45.22 &   0.97 &      0.41 &   2.74 &      1.66 &   4.60 &      2.11 &    7.73 &       2.75 \\
3 &  15:34:35.329 & -00:28:45.74 &   0.23 &      0.10 &   0.57 &      0.33 &   0.88 &      0.25 &    1.89 &       0.27 \\
4 &  15:34:35.406 & -00:28:46.04 &   0.64 &      0.49 &   1.68 &      1.27 &   3.74 &      1.94 &    6.01 &       2.52 \\
5 &  15:34:35.422 & -00:28:45.23 &   0.01 &      0.01 &   0.59 &      0.46 &   7.34 &      6.52 &   15.78 &       5.73 \\
6 &  15:34:35.454 & -00:28:45.14 &   0.18 &      0.13 &   0.68 &      0.36 &   1.81 &      2.32 &    3.58 &       2.13 \\
7 &  15:34:35.351 & -00:28:44.40 &   0.27 &      0.13 &   1.04 &      0.54 &   2.98 &      1.54 &    6.75 &       5.55 \\
8 &  15:34:35.384 & -00:28:43.83 &   0.70 &      0.28 &   2.34 &      1.04 &   8.28 &      4.04 &   16.27 &       5.96 \\
9 &  15:34:35.344 &-00:28:43.19  &   0.31 &      0.10 &   1.11 &      0.56 &   2.89 &      1.24 &    6.15 &       2.15 \\
10&  15:34:35.440 &-00:28:43.54  &   0.96 &      0.36 &   2.88 &      1.67 &   6.04 &      8.46 &   11.20 &       9.74 \\
11&  15:34:35.453 &-00:28:43.46  &   0.68 &      0.39 &   2.03 &      1.18 &   3.80 &      2.96 &    6.92 &       2.96 \\
12&  15:34:35.374 &-00:28:42.74  &   0.95 &      0.63 &   2.96 &      2.27 &   5.51 &      3.36 &    8.75 &       4.90 \\
   \bottomrule
   	\toprule
     G04-1 \\
   	\hline
1&  04:12:19.749  &  -05:54:47.13  &  1.42 &      1.00 &   2.50 &      1.84 &   3.66 &      2.15 &    5.10 &       1.84 \\
2&  04:12:19.758  &  -05:54:47.73  &  2.09 &      0.86 &   4.39 &      1.83 &  10.33 &      6.18 &   17.55 &      10.33 \\
3&  04:12:19.773  &  -05:54:48.04  &  1.13 &      0.61 &   2.31 &      1.18 &   5.06 &      8.30 &    9.21 &      11.04 \\
4&  04:12:19.789  &  -05:54:48.37  &  0.68 &      0.18 &   1.54 &      0.68 &   4.37 &     11.51 &    7.26 &      12.25 \\
5&  04:12:19.805  &  -05:54:48.99  &  0.71 &      0.32 &   1.27 &      0.57 &   3.51 &      3.53 &    6.38 &       2.94 \\
6&  04:12:19.781  &  -05:54:49.92  &  1.67 &      1.37 &   2.52 &      1.92 &   4.57 &      4.97 &    6.55 &       4.59 \\
7&  04:12:19.760  &  -05:54:50.42  &  0.43 &      0.23 &   0.67 &      0.45 &   1.32 &      0.67 &    1.76 &       1.27 \\
8&  04:12:19.740  &  -05:54:48.56  &  3.89 &      2.55 &   8.06 &      4.85 &  17.73 &     21.53 &   30.53 &      25.08 \\
9&  04:12:19.704  &  -05:54:48.81  &  4.38 &      3.07 &   8.74 &      5.32 &  21.67 &     15.65 &   40.43 &      23.87 \\
10&  04:12:19.698  &  -05:54:48.34 &  5.46 &      2.18 &  12.86 &      7.31 &  35.11 &     14.73 &   80.70 &      24.93 \\
11&  04:12:19.647  &  -05:54:48.56 &  1.29 &      0.86 &   2.62 &      1.18 &   6.87 &      4.37 &   11.63 &      10.81 \\
12&  04:12:19.722  &  -05:54:49.70 &  1.10 &      0.60 &   2.27 &      0.96 &   5.79 &      1.70 &   10.21 &       5.13 \\
13&  04:12:19.648  &  -05:54:49.27 &  0.24 &      0.03 &   0.88 &      0.37 &   2.17 &     13.44 &    4.52 &       3.70 \\
14&  4:12:19.7189 &  -05:54:47.289 &  1.29 &      0.94 &   2.51 &      1.80 &   4.07 &      1.91 &    5.82 &       2.17 \\
    \bottomrule
  	 \toprule
     G20-2 \\
  	\hline
1 & 20:44:02.988  &  -06:46:56.35     &   0.75 &      0.40 &   1.28 &      0.78 &   1.64 &      0.75 &    2.97 &       1.25 \\
2 & 20:44:03.013  &  -06:46:57.61     &   1.16 &      0.65 &   1.65 &      0.86 &   2.59 &      1.79 &    3.54 &       1.77 \\
3 & 20:44:02.965  &  -06:46:57.34     &   5.13 &      1.55 &   9.71 &      3.99 &  17.73 &      7.22 &   26.06 &       8.04 \\
4 & 20:44:03.020  &  -06:46:58.08     &   0.57 &      0.46 &   0.98 &      0.51 &   2.82 &      2.92 &    2.97 &       2.12 \\
5 & 20:44:02.950  &  -06:46:57.49     &   11.63 &      3.31 &  23.42 &     13.94 &  56.73 &     34.76 &   89.36 &      49.92 \\
6 & 20:44:02.959  &  -06:46:58.07     &   6.28 &      2.67 &  11.96 &      5.55 &  31.53 &     13.20 &   50.04 &      21.77 \\
7 & 20:44:02.984  &  -06:46:58.60     &   2.14 &      1.77 &   2.60 &      2.22 &   4.24 &      3.07 &    5.02 &       3.54 \\
8 & 20:44:02.925  &  -06:46:57.83     &   2.90 &      0.84 &   7.12 &      3.14 &  28.45 &     27.90 &   50.83 &      42.64 \\
9 &  20:44:02.944 &  -06:46:58.30    	&   7.60 &      2.91 &  13.51 &      5.50 &  35.84 &     12.99 &   53.89 &      19.14 \\
10&  20:44:02.923 &  -06:46:58.44     &   4.50 &      3.07 &   7.64 &      5.07 &  18.70 &     10.28 &   25.78 &      13.19 \\
11&  20:44:02.906 &  -06:46:58.25     &   6.49 &      4.61 &  12.02 &      6.22 &  32.02 &     16.14 &   47.34 &      21.14 \\
12&  20:44:02.895 &  -06:46:57.80     &  18.02 &     10.27 &  36.41 &     27.26 &  99.79 &     53.69 &  155.33 &      65.25 \\
  \bottomrule
  	\toprule
    G08-5 \\
  	\hline
1& 8:54:18.829  &  +6:46:20.679&   0.33 &      0.20 &   0.63 &      0.36 &   1.04 &      0.87 &    2.12 &       0.97 \\
2& 8:54:18.820  &  +6:46:20.533&   0.45 &      0.46 &   0.88 &      0.35 &   2.15 &      1.57 &    4.21 &       4.77 \\
3& 8:54:18.788	 & +6:46:19.895 &   0.44 &      0.16 &   2.05 &      1.44 &   5.05 &      3.59 &   10.26 &       6.91 \\
4& 8:54:18.771	 & +6:46:19.911  &   0.78 &      0.32 &   2.96 &      1.69 &   7.21 &      3.30 &   15.13 &      10.86 \\
5& 8:54:18.711	 & +6:46:19.227  &   0.70 &      0.67 &   1.36 &      1.12 &   2.33 &      1.57 &    2.98 &       1.43 \\
6& 8:54:18.795	 & +6:46:20.936  &   2.36 &      1.79 &   3.61 &      2.45 &   6.39 &      5.10 &    8.25 &       4.19 \\
7& 8:54:18.790	 & +6:46:21.158 &   4.55 &      3.26 &   6.85 &      4.58 &  11.12 &      6.82 &   14.47 &       8.17 \\
8& 8:54:18.794	 & +6:46:21.328  &   4.35 &      3.18 &   6.55 &      5.10 &  10.63 &      7.03 &   13.86 &       8.40 \\
    \bottomrule
  	\toprule
  	\hline
  	G14-1 \\
  	\hline
1 &14:54:28.387 &+00:44:34.27&   2.58 &      2.17 &   6.11 &      4.90 &  12.84 &     11.97 &   23.33 &      18.73 \\
2 &14:54:28.324 &+00:44:34.59&  13.71 &     11.80 &  25.17 &     23.21 &  39.50 &     30.96 &   55.99 &      42.53 \\
3 &14:54:28.307 & +00:44:33.91&  3.17 &      2.82 &   5.23 &      4.07 &  10.09 &      7.92 &   13.19 &       7.61 \\

\bottomrule
\end{longtable}
\end{center}

\begin{center}
\onecolumn
\begin{longtable}{lcccc}
\caption{Properties of clumps in DYNAMO galaxies}\\
\toprule
\label{tab:mass_measurements}
 ID& $\log{M}_{\star,raw}$&$\log {M}_{\star,disksub}$ & $Position_{clump}$/$R_{gal,85\%}$ & Size \\ D13-5  & [$\mathrm{M}_\odot$]&[$\mathrm{M}_\odot$]& &[Kpc]\\
\midrule
1 &         7.14 &             6.90 &         0.53 &      0.25 \\
2 &         7.48 &             7.32 &         0.72 &      0.34 \\
3 &         7.68 &             7.20 &         0.37 &      0.29 \\
4 &         7.21 &             7.13 &         0.34 &      0.27 \\
5 &         7.92 &             6.95 &         0.34 &      0.55 \\
6 &         7.75 &             7.37 &         0.34 &      0.46 \\
7 &         7.34 &             7.21 &         0.40 &      0.24 \\
8 &         8.03 &             7.44 &         0.27 &      0.39 \\
9 &         7.95 &             7.63 &         0.34 &      0.39 \\
10 &         7.76 &             6.39 &         0.35 &      0.43 \\
11 &         7.94 &             7.77 &         0.34 &      0.44 \\
12 &         7.23 &             6.00 &         0.46 &      0.19 \\
13 &         7.78 &             7.80 &         0.34 &      0.32 \\
14 &         7.59 &             7.02 &         0.89 &      0.25 \\
15 &         6.94 &             6.58 &         0.83 &      0.28 \\
16 &         7.75 &             5.65 &         0.95 &      0.50 \\
17 &         6.87 &             6.37 &         1.17 &      0.22 \\
\bottomrule
    \toprule
     D15-3 \\
  	\hline
1 &         7.52 &             7.06 &         0.47 &      0.28 \\
2 &         7.17 &             6.80 &         0.68 &      0.24 \\
3 &         6.67 &             5.47 &         0.47 &      0.15 \\
4 &         7.35 &             6.82 &         0.48 &      0.26 \\
5 &         7.71 &             7.20 &         0.29 &      0.28 \\
6 &         7.21 &             7.14 &         0.35 &      0.21 \\
7 &         7.57 &             7.41 &         0.18 &      0.22 \\
8 &         8.11 &             7.68 &         0.13 &      0.34 \\
9 &         7.43 &             6.78 &         0.36 &      0.25 \\
10 &         7.62 &             7.88 &         0.28 &      0.34 \\
11 &         7.33 &             7.11 &         0.35 &      0.28 \\
12 &         7.48 &             7.01 &         0.45 &      0.26 \\

\bottomrule
    \toprule
     G04-1 \\
  	\hline

1 &         7.35 &             6.80 &         0.68 &      0.41 \\
2 &         8.35 &             8.47 &         0.50 &      0.65 \\
3 &         7.82 &             8.39 &         0.49 &      0.48 \\
4 &         8.08 &             8.80 &         0.56 &      0.54 \\
5 &         8.18 &             7.77 &         0.70 &      0.65 \\
6 &         7.68 &             7.70 &         0.89 &      0.60 \\
7 &         7.22 &             6.99 &         1.02 &      0.40 \\
8 &         8.58 &             8.62 &         0.22 &      0.59 \\
9 &         8.78 &             8.72 &         0.14 &      0.62 \\
10 &         9.01 &             8.63 &         0.06 &      0.77 \\
11 &         8.26 &             8.48 &         0.48 &      0.52 \\
12 &         8.19 &             7.61 &         0.61 &      0.64 \\
13 &         7.75 &             8.28 &         0.62 &      0.43 \\
14 &         7.66 &             6.91 &         0.58 &      0.39 \\
\bottomrule
    \toprule
     G20-2\\
  	\hline
1 &         7.27 &             6.61 &         1.16 &      0.60 \\
 2 &         7.42 &             7.30 &         0.98 &      0.51 \\
 3 &         8.40 &             8.02 &         0.47 &      0.79 \\
 4 &         7.70 &             7.53 &         1.18 &      0.59 \\
 5 &         9.25 &             9.00 &         0.40 &      1.12 \\
 6 &         9.09 &             8.53 &         0.58 &      0.87 \\
 7 &         7.72 &             7.24 &         0.99 &      0.47 \\
 8 &         9.23 &             9.31 &         0.13 &      0.52 \\
 9 &         8.99 &             8.59 &         0.55 &      0.92 \\
10 &         8.63 &             8.20 &         0.55 &      0.76 \\
11 &         8.93 &             8.65 &         0.48 &      0.89 \\
12 &        9.57 &              9.032&         0.30 &       1.44\\
\bottomrule
    \toprule
     G08-5 \\
  	\hline
1 &         7.39 &             7.05 &         0.66 &      0.42 \\
2 &         7.80 &             8.03 &         0.66 &      0.52 \\
3 &         8.08 &             8.31 &         0.39 &      0.54 \\
4 &         8.26 &             8.35 &         0.35 &      0.69 \\
5 &         7.41 &             6.88 &         0.79 &      0.43 \\
6 &         7.79 &             7.58 &         0.46 &      0.54 \\
7 &         8.01 &             7.69 &         0.55 &      0.71 \\
8 &         8.00 &             7.71 &         0.64 &      0.73 \\
\bottomrule
    \toprule
     G14-1 \\
  	\hline
1 &         8.24 &             8.45 &         0.33 &      0.57 \\
2 &         8.58 &             8.26 &         0.45 &      0.44 \\
3 &         8.05 &             7.89 &         0.63 &      0.64 \\
\bottomrule

\end{longtable}
\end{center}
\newpage
\bsp	
\label{lastpage}
\end{document}